\documentstyle[aps,tighten,eqsecnum,graphics]{revtex} 

\begin{document}

\title{Limits to squeezing in the degenerate OPO}

\author{S.~Chaturvedi~\( ^{1} \), K.~Dechoum~\( ^{2} \) and P.~D.~Drummond~\( ^{2} \) }

\address{ \( ^{1} \) School of Physics, University of Hyderabad, Hyderabad
500046, India\\
 \( ^{2} \) Department of Physics, University of Queensland, St
Lucia 4067, Queensland, Australia 
}

\maketitle

\begin{abstract}
We develop a systematic theory of quantum fluctuations in the driven parametric oscillator (OPO), including the region near threshold. This allows us to treat the limits imposed by nonlinearities to quantum squeezing and noise reduction, in this non-equilibrium quantum phase-transition. In particular, we compute the squeezing spectrum near threshold, and calculate the optimum value. We find that the optimal noise reduction occurs at different driving fields, depending on the ratio of damping rates. The largest spectral noise reductions are predicted to occur with a very high-Q second-harmonic cavity. Our analytic results agree well with stochastic numerical simulations. We also compare the results obtained in the positive-P representation, as a fully quantum mechanical calculation, with the truncated Wigner phase space equation, also known as semiclassical theory.  
\end{abstract}

\section{Introduction}

Optical parametric oscillators are one of the most interesting and well characterized devices in nonlinear quantum optics. Novel discoveries made with them include demonstrations of large amounts of squeezing\cite{WKHW86}, significant quantum intensity correlations\cite{HHRG87} together with a quadrature correlation measurement that provided the first experimental demonstration \cite{OPK92} of the original EPR paradox. Practical applications include their use as highly efficient and tunable frequency converters. In the present paper, we focus on the optimum below-threshold squeezing results, which determine the limits to squeezing obtained near the critical point, where nonlinear corrections start to dominate. In a companion paper\cite{Drummond_crit}, the related question of critical fluctuations at threshold is treated.

The theory of quantum squeezing in the linear parametric oscillator is now well-developed\cite{Drummond.et.al}-\cite{Veits}. Excellent agreement between theory and experiment is obtained\cite{WKHW86}-\cite{HHRG87}, in the region below threshold. However, the usual theory is linearized, and therefore cannot be used in the near-threshold region where the squeezing is largest. The drawback with linearized theories is that they predict that zero quantum noise levels are achievable at threshold. This is clearly unrealistic, since (by the Heisenberg uncertainty principle) it necessarily requires an infinite energy in the conjugate mode. More significantly, this would imply an infinite amount of phase information - which is also impossible, since the coherent pump which drives the parametric oscillator can only supply a finite quantity of phase information.

While present experiments are limited by technical noise from approaching the critical point too closely, it is reasonable to expect that progress in integrated optics will lead to more stable, highly miniaturized devices which could well operate at the quantum limit, even near threshold. Accordingly, there have been a number of investigations as to the ultimate limits to the squeezing spectrum of a parametric amplifier/oscillator. This has often involved using Keldysh diagrams or Wyld-Keldysh techniques \cite{Schwinger}-\cite{Wyld} to extend the linear theory \cite{Kennedy}-\cite{Veits}, using a many-body theory analog of Feynman diagrams.

The two-mode Hilbert space involved in these problems typically has a minimum dimension \( >10^{6} \), even with only \( N=10^{3} \) photons , and therefore would be difficult to solve using other methods that involve number-state expansions - either using a direct solution of the master equation, or stochastic wave-function methods. The Hamiltonian matrix would have \( 10^{12} \) coefficients, unless simplified, with a density matrix of similar size. More typical experimental photon numbers have at least \( N=10^{6} \), with a corresponding density matrix dimension of \( 10^{24} \) - which appears completely inaccessible with number state techniques. Another drawback of number-state techniques is that they usually do not permit analytic approximations, which can give more physical insight.

We therefore treat these questions using the coherent-state positive P-representation \cite{+P}, combined with an expansion technique valid below the critical point. Results are also verified by the use of direct numerical stochastic equation simulations. We find an \( N^{-2/3} \) scaling law for the optimal squeezing predicted by Plimak and Walls\cite{Plimak} is obtained here as well, but with a different spectrum, owing to the use of more systematic expansion techniques that result from using the positive P-representation method. Our analytic results for optimal squeezing, which occurs below the critical threshold, give excellent agreement with accurate numerical simulations for the same parameter values. However, even larger noise reductions are predicted to occur simply by reducing the losses of the second-harmonic, in which case the \( N^{-2/3} \) scaling law no longer holds. In a companion paper we consider the related problem of the critical region, where the narrow-band squeezing is less than optimal due to the effects of critical fluctuations.

We also compare the above results with a semiclassical approach, that is, a truncated Wigner phase-space equation. This equation corresponds to a classical theory with added vacuum fluctuations. A comparison between the positive P-representation (fully quantum mechanical) and semiclassical theories permits us to see how far one can go and what is the limitation of this extended classical point of view. We find that the nonlinear corrections in the semiclassical theory are in strong disagreement with the full quantum theory far below threshold, but agree near threshold. This tells us that the semiclassical theory works surprisingly well in the threshold region, indicating that the large quantum fluctuations near threshold have a rather classical character.

\section{Hamiltonian and master equation}

The model considered here is the degenerate parametric oscillator. The system of interest is an idealized interferometer, which is resonant at two frequencies, \( \omega _{1} \) and \( \omega _{2}\approx 2\omega _{1} \). It is externally driven at the larger of the two frequencies. Both frequencies are damped due to cavity losses. Down conversion of the pump photons to resonant sub-harmonic mode photons occurs due to a \( \chi ^{(2)} \) nonlinearity present inside the cavity. The Heisenberg picture Hamiltonian that describes this open system\cite{Drummond.et.al} is \begin{equation}
\label{a1}
\widehat{H}=\widehat{H}_{sys}+\sum _{j=1,2}\hbar \left( \widehat{a}_{j}\widehat{\Gamma }_{j}^{\dagger }+\widehat{a}_{j}^{\dagger }\widehat{\Gamma }_{j}\right) +\widehat{H}_{R}\, \, ,
\end{equation}
 where the intra-cavity or system Hamiltonian is given by: \begin{equation}
\label{a2}
\widehat{H}_{sys}=\sum _{j=1,2}\hbar \omega _{j}\widehat{a}_{j}^{\dagger }\widehat{a}_{j}+i\hbar \frac{\chi }{2}\left( \widehat{a}_{1}^{\dagger 2}\widehat{a}_{2}-\widehat{a}_{1}^{2}\widehat{a}_{2}^{\dagger }\right) +\, \, i\hbar \left( {\mathcal{E}}e^{-i\omega _{2}t}\widehat{a}_{2}^{\dagger }-{\mathcal{E}}^{\ast }e^{i\omega _{2}t}\widehat{a}_{2}\right) \, \, .
\end{equation}

Here \( \mathcal{E} \) represents the external driving field at frequency \( \omega _{2} \). The term \( \widehat{H}_{R} \) describes the free evolution of the extra-cavity modes that are the loss-reservoirs of the cavity. The term \( \chi  \) is the coupling parameter due to a \( \chi ^{(2)} \) nonlinear medium internal to the cavity, and \( \widehat{\Gamma }_{j}^{\dagger },\Gamma _{j} \) are reservoir operators that create and destroy photons in the loss-reservoir coupled to the internal mode of frequency \( \omega _{j} \) .

Next, we wish to consider an interaction picture, obtained with the definition that \begin{equation}
\widehat{H}_{0}=\sum _{j=1,2}\hbar \omega _{j}\widehat{a}_{j}^{\dagger }\widehat{a}_{j}\, \, .\end{equation}
 In other words, the operators will evolve according to the relevant mode frequency, while the states evolve according to the rest of the system Hamiltonian. The interaction Hamiltonian used here then reduces to the the standard one\cite{Drummond.et.al} for a non-degenerate, single-mode parametric amplifier or oscillator: \begin{equation}
\label{a4}
\widehat{H}_{int}/\hbar =i{\mathcal{E}}\left[ \widehat{a}_{2}-\widehat{a}_{2}^{\dagger }\right] +\frac{i\chi }{2}\left[ \widehat{a}_{2}\widehat{a}_{1}^{\dagger2 }-\widehat{a}_{2}^{\dagger }\widehat{a}_{1}^{2}\right] \, \, .
\end{equation}
 Here \( \widehat{a}_{1},\widehat{a}_{2} \) are now time-independent operators representing the fundamental and second-harmonic modes respectively. For simplicity, we have chosen the field mode-functions so that \( \mathcal{E} \), \( \chi  \) are real.

Using standard techniques \cite{HJC_Book} to eliminate the heat bath, we obtain the following master equation for the reduced density operator of the system in the interaction picture: \begin{eqnarray}
\frac{\partial \widehat{\rho }}{\partial t} & = & \frac{1}{i\hbar }\left[ \widehat{H}_{int},\widehat{\rho }\right] +\gamma _{1}\left( 2\widehat{a}_{1}\widehat{\rho }\widehat{a}_{1}^{\dagger }-\widehat{a}_{1}^{\dagger }\widehat{a}_{1}\widehat{\rho }-\widehat{\rho }\widehat{a}_{1}^{\dagger }\widehat{a}_{1}\right) \nonumber \\
 & + & \gamma _{2}\left( 2\widehat{a}_{2}\widehat{\rho }\widehat{a}_{2}^{\dagger }-\widehat{a}_{2}^{\dagger }\widehat{a}_{2}\widehat{\rho }-\widehat{\rho }\widehat{a}_{2}^{\dagger }\widehat{a}_{2}\right) \, \, ,\label{a5} 
\end{eqnarray}

\noindent where \( \gamma _{i} \) are the internal mode amplitude damping rates and we assume that \( {\bar{n}}_{i}<<1 \) , where \( {\bar{n}}_{i} \) are the mean numbers of thermal photons in the input reservoir modes. Using reservoir theory, it is possible to identify the coherent driving field with a corresponding input photon flux from an external coherent laser, with \( I_{2}=|{\mathcal{E}}|^{2}/2\gamma ^{in}_{2} \) in \( photons/s \) , where \( \gamma ^{in}_{2} \) is the input coupler decay rate. For optimum performance, we will assume that \( \gamma _{2}=\gamma ^{in}_{2} \) , and similarly for the fundamental mode - which will be assumed to only decay through its output coupling mirror. If these conditions are not satisfied, then the coupling efficiency and maximum squeezing are reduced.

At this point, we note that in the classical limit, the system has the well-known classical equations of intra-cavity parametric oscillation, where we define: \( \alpha _{i}=\langle \widehat{a}_{i}\rangle  \), and hence obtain, in the interaction picture:

\begin{eqnarray}
\frac{d\alpha _{1}}{dt} & = & \left[ -\gamma _{1}\alpha _{1}+\chi \alpha _{1}^{*}\alpha _{2}\right] \, ,\nonumber \\
\frac{d\alpha _{2}}{dt} & = & \left[ -\gamma _{2}\alpha _{2}+{\mathcal{E}}-\frac{1}{2}\chi \alpha _{1}^{2}\right] 
\end{eqnarray}
 These equations are valid in the limit of large photon number. They are obtained by the use of a classical decorrelation in which all operator products are assumed to factorize, so that \( \langle \widehat{a}_{i}^{\dagger }\widehat{a}_{j}\rangle \simeq \langle \widehat{a}_{i}^{\dagger }\rangle \langle \widehat{a}_{j}\rangle  \), and \( \langle \widehat{a}_{i}\widehat{a}_{j}\rangle \simeq \langle \widehat{a}_{i}\rangle \langle \widehat{a}_{j}\rangle  \). The solution of these equations is immediate classically, and has the property that there is a phase transition at the critical driving field of: \( {\mathcal{E}}={\mathcal{E}}_{c}=\gamma _{1}\gamma _{2}/\chi  \) . For driving fields below this value, one has: \begin{eqnarray}
\alpha _{1} & = & 0\, ,\nonumber \\
\alpha _{2} & = & {\mathcal{E}}/\gamma _{2}\, .
\end{eqnarray}
For fields above this value, the signal field \( \alpha _{1} \) is bistable, with:

\begin{eqnarray}
\alpha _{1} & = & \pm \sqrt{\frac{2}{\chi }\left( {\mathcal{E}}-{\mathcal{E}}_{c}\right) }\, ,\nonumber \\
\alpha _{2} & = & \frac{\gamma _{1}}{\chi }\, .
\end{eqnarray}

The intra-cavity photon number at the critical point is \( N_{c}=\gamma _{1}^{2}/\chi ^{2}={\mathcal{E}}_{c}^{2}/\gamma _{2}^{2} \) . Classically, there are only second-harmonic photons present at this driving field, and the input photon flux is \( I_{c}={\mathcal{E}}_{c}^{2}/2\gamma _{2}=\gamma _{1}^{2}\gamma _{2}/4\chi ^{2} \). However, a squeezed field - with finite intensity - is actually emitted as well. This is not taken into account in the classical theory.

\section{Operator Representations}

In order to treat the full quantum evolution, we now turn to the methods of operator representation theory. These techniques can be used to transform the density matrix equations of motion to c-number Fokker-Planck or stochastic equations.

\subsection{The positive P-representation}

In the positive P-representation,  the density matrix is expanded 
in terms of multi-mode coherent state vectors \( |\overrightarrow{\alpha }\rangle  \):
\begin{equation}
\label{a6}
\widehat{\rho }=\int P\left( \overrightarrow{\alpha },\overrightarrow{\alpha }^{+}\right) \frac{|\overrightarrow{\alpha }\rangle \langle \left( \overrightarrow{\alpha }^{+}\right) ^{\ast }|}{\langle \left( \overrightarrow{\alpha }^{+}\right) ^{\ast }|\overrightarrow{\alpha }\rangle }d^4 \overrightarrow{\alpha }d^4\overrightarrow{\alpha }^{+} \, \, .
\end{equation}

Following standard procedures, the assumption of vanishing boundary terms allows the master equation to be re-written as a Fokker-Planck equation in \( P\left( \overrightarrow{\alpha }\overrightarrow{,\alpha }^{+}\right) , \) and hence as a stochastic equation\cite{Arnold} with real noise. The assumption of vanishing boundary terms is critical to this procedure, and we note here that this is generally valid when the ratio of nonlinearity to damping is small\cite{Boundary}, (i.e. \( \left| \chi /\gamma _{k}\right| \ll1  \)). The stochastic procedure is best regarded as being generally an asymptotic procedure, valid for small \( \left| \chi /\gamma _{k}\right|  \) - in which case the boundary terms are exponentially suppressed. We check this assumption numerically here as well, and point out that the required ratio of nonlinearity to damping is extremely well-satisfied in current experiments, where the ratio is typically 10\( ^{-6} \) or less. Further analysis of this problem has been given elsewhere\cite{Boundary}. Given this assumption, the following stochastic equations are obtained from (\ref{a5}) and (\ref{a6}), for any driving field \( \mathcal{E} \), that is, either below or above threshold: \begin{eqnarray}
d\alpha _{1} & = & \left[ -\gamma _{1}\alpha _{1}+\chi \alpha _{1}^{+}\alpha _{2}\right] dt+\sqrt{\chi \alpha _{2}}dw_{1}(t)\, \, ,\nonumber \\
d\alpha _{1}^{+} & = & \left[ -\gamma _{1}\alpha ^{+}_{1}+\chi \alpha _{1}\alpha _{2}^{+}\right] dt+\sqrt{\chi \alpha _{2}^{+}}dw_{2}(t)\, \, ,\nonumber \\
d\alpha _{2} & = & \left[ -\gamma _{2}\alpha _{2}+{\mathcal{E}}-\frac{1}{2}\chi \alpha _{1}^{2}\right] dt\, \, ,\nonumber \\
d\alpha _{2}^{+} & = & \left[ -\gamma _{2}\alpha _{2}^{+}+{\mathcal{E}}-\frac{1}{2}\chi \alpha _{1}^{+2}\right] dt\, \, .
\end{eqnarray}
 The stochastic correlations are given by: \begin{eqnarray}
\left\langle dw_{k}(t)\right\rangle  & = & 0\, \, ,\nonumber \\
\left\langle dw_{k}(t)dw_{l}(t)\right\rangle  & = & \delta _{kl}dt\, \, .
\end{eqnarray}

This means that \( dw_{k}(t) \) represent two real Gaussian and uncorrelated stochastic processes, and the amplitude of the stochastic fluctuations that act on the signal mode are dependent on the pump field dynamics. Our derivation is formally based on the It\^{o} stochastic calculus. However, in this case, either It\^{o} or Stratonovic stochastic calculus gives identical results\cite{Arnold}.

\subsection{The semiclassical theory}

\noindent We can also write a c-number phase space equation using an approximate form of the Wigner representation\cite{Kinsler-Drummond91}, which is equivalent to stochastic electrodynamics. The characteristic function of the Wigner representation is written as \begin{equation}
\chi _{W}(z)=Tr\left( \rho e^{iz^{*}\widehat{a}^{\dagger }+iz\widehat{a}}\right) =Tr\left( \rho e^{iz^{*}\widehat{a}^{\dagger }}e^{iz\widehat{a}}e^{-|z|^{2}/2}\right) \, \, ,
\end{equation}
 and the Wigner distribution can be written as Fourier transform of the characteristic function \begin{equation}
W(\alpha )=\frac{1}{\pi ^{2}}\int _{-\infty }^{\infty }d^{2}z\chi _{W}(z)e^{-iz^{*}\alpha ^{*}}e^{-iz\alpha }\, \, .
\end{equation}
 In the Wigner representation, the phase space equation that corresponds to the master equation (\ref{a5}) is \begin{eqnarray}
\frac{\partial W(\alpha _{1},\alpha _{2},t)}{\partial t} & = & \left\{ \frac{\partial }{\partial \alpha _{1}}\left( \gamma _{1}\alpha _{1}-\chi \alpha _{1}^{*}\alpha _{2}\right) +\frac{\partial }{\partial \alpha _{1}^{*}}\left( \gamma _{1}\alpha _{1}^{*}-\chi \alpha _{1}\alpha _{2}^{*}\right) \right. \nonumber \\
 & + & \frac{\partial }{\partial \alpha _{2}}\left( \gamma _{2}\alpha _{2}+\frac{\chi }{2}\alpha _{1}^{2}-{\mathcal{E}}\right) +\frac{\partial }{\partial \alpha _{2}^{*}}\left( \gamma _{2}\alpha _{2}^{*}+\frac{\chi }{2}\alpha _{1}^{*2}-{\mathcal{E}}\right) \nonumber \\
 & + & \gamma _{1}\left( 1+2{\bar{n}}_{1}\right) \frac{\partial ^{2}}{\partial \alpha _{1}\partial \alpha _{1}^{*}}+\gamma _{2}\left( 1+2{\bar{n}}_{2}\right) \frac{\partial ^{2}}{\partial \alpha _{2}\partial \alpha _{2}^{*}}\nonumber \\
 & + & \left. \frac{\chi }{8}\left( \frac{\partial ^{3}}{\partial \alpha _{1}^{*2}\partial \alpha _{2}}+\frac{\partial ^{3}}{\partial \alpha _{1}^{2}\partial \alpha _{2}^{*}}\right) \right\} W(\alpha _{1},\alpha _{2},t)\, \, .
\end{eqnarray}

If we truncate the third derivative of the phase space equation we get a genuine Fokker-Planck type equation with positive definite diffusion constant. This can be mapped into the following It\^{o} stochastic differential coupled equations (for simplicity we let \( {\bar{n}}_{i}=0 \), as before). \begin{eqnarray}
d{\alpha }_{1} & = & \left[ -\gamma _{1}{\alpha }_{1}+\chi \alpha _{1}^{*}{\alpha }_{2}\right] dt+\sqrt{\gamma _{1}}\; dw_{1}(t)\, \, ,\nonumber \\
d\alpha _{1}^{*} & = & \left[ -\gamma _{1}\alpha _{1}^{*}+\chi {\alpha }_{1}\alpha _{2}^{*}\right] dt+\sqrt{\gamma _{1}}\; dw_{1}^{*}(t)\, \, ,\nonumber \\
d{\alpha }_{2} & = & \left[ -\gamma _{2}{\alpha }_{2}-\frac{\chi }{2}{\alpha }_{1}^{2}+{\mathcal{E}}\right] dt+\sqrt{\gamma _{2}}\; dw_{2}(t)\, \, ,\nonumber \\
d\alpha _{2}^{*} & = & \left[ -\gamma _{2}\alpha _{2}^{*}-\frac{\chi }{2}\alpha _{1}^{*2}+{\mathcal{E}}\right] dt+\sqrt{\gamma _{2}}\; dw_{2}^{*}(t)\, \, .
\end{eqnarray}

\noindent Here \( dw_{k}(t) \) is now a \emph{complex} Gaussian white noise whose mean and variance are given by \begin{eqnarray}
\langle dw_{k}(t)\rangle  & = & 0\; \; \; ,\nonumber \\
\langle dw_{k}(t)dw_{l}^{*}(t)\rangle  & = & \delta _{kl}dt\, \, .\label{a8} 
\end{eqnarray}

The above equation is identical to the equation derived in positive P-representation when one discards the noise terms. This corresponds to the nonlinear classical equation for the OPO system. The main difference between the two sets of equations is the noise terms. In the semiclassical theory the noise is universal for all modes and comes from to the vacuum fluctuations, while in the positive-P equation the pump has a noiseless amplitude and the signal noise comes from the nonlinear coupling. However, we note that the Wigner equation after truncation is no longer completely equivalent to quantum mechanics, since it always leads to a positive Wigner function - thus, not all quantum states can be represented.

\subsection{Observable moments and spectra}

The positive-P stochastic method directly reproduces the normally ordered correlations and moments, while the Wigner representation reproduces the symmetrically ordered moments. We also have to distinguish the internal and external operator moments, since measurements are normally performed on output fields that are external to the cavity. The technique for treating external field spectra was introduced by Yurke\cite{Yurke}, and by Collett and Gardiner\cite{Gardiner}.

These external field measurements are obtained from the input-output relations of: \begin{equation}
\widehat{\Phi }^{out}_{j}(t)=\sqrt{2\gamma ^{out}_{j}}\widehat{a}_{j}(t)-\widehat{\Phi }^{in}_{j}(t)\, \, ,
\end{equation}
 where \( \widehat{\Phi }^{in}_{j}(t) \) and \( \widehat{\Phi }^{out}_{j}(t) \) are the input and output photon fields respectively, evaluated at the output-coupling mirror. The most efficient transport of squeezing is obtained if we assume that all the signal losses occur through the output coupler, so that \( \gamma _{1}=\gamma _{1}^{out} \). We will assume this to be the case.

The crucial quadrature variables of the system have the definitions: \begin{eqnarray}
\widehat{x}_{j} & = & \left( \widehat{a}_{j}+\widehat{a}_{j}^{\dagger }\right) \, \, ,\nonumber \\
\widehat{y}_{j} & = & \frac{1}{i}\left( \widehat{a}_{j}-\widehat{a}_{j}^{\dagger }\right) \, \, .\label{quadr} 
\end{eqnarray}

There are also corresponding external quadrature field variables, defined as:

\begin{eqnarray}
\widehat{X}_{j} & = & \left( \widehat{\Phi }^{out}_{j}+\widehat{\Phi }^{out\dagger }_{j}\right) \, \, ,\nonumber \\
\widehat{Y}_{j} & = & \frac{1}{i}\left( \widehat{\Phi }^{out}_{j}-\widehat{\Phi }^{out\dagger }_{j}\right) \, \, .\label{quadrX} 
\end{eqnarray}

Similarly, we can define c-number stochastic quadrature variables within the relevant representations, thus giving:

\begin{eqnarray}
{x}_{j} & = & \left( \alpha _{j}+\alpha _{j}^{+}\right) \, \, ,\nonumber \\
y_{j} & = & \frac{1}{i}\left( \alpha _{j}-\alpha _{j}^{+}\right) \, \, \label{quadrx} 
\end{eqnarray}

Of especial interest is \( {\widehat{Y}}_{1} \) since this is the low-noise, squeezed quadrature. Here we note that the instantaneous correlation functions of the intra-cavity field operators are called the moments. Typically, they are not easily measurable, when compared to output moments or spectra, but they are useful in that they provide a check on the accuracy of the calculation of measurable spectra.

The squeezing in terms of the intra-cavity quadrature variances corresponds to an instantaneous measurement of the field moments. If such a measurements were possible, it would include contributions from all frequencies. For measurements averaged over a long time \( T \), it is the low frequency part of the spectrum that is the relevant quantity, and we shall focus on this, as it usually determines the maximum squeezing possible. The output measured spectral variance \( V^{\theta }_{j} \) of a general quadrature \[
\widehat{X}_{j}^{\theta }=\left( e^{-i\theta }\widehat{\Phi }_{j}^{out}+e^{i\theta }\widehat{\Phi }_{j}^{out\dagger }\right) \, \, ,\]
 can be written

\begin{equation}
\label{OS-spectra}
V^{\theta }_{j}(\omega )\delta (\omega +\omega ^{\prime })=\left< \Delta \widehat{X}_{j}^{\theta }(\omega )\Delta \widehat{X}_{j}^{\theta }(\omega ^{\prime })\right> \, \, .
\end{equation}
 where the fluctuations \( \Delta \widehat{X}_{j}^{\theta } \) are defined as \( \Delta \widehat{X}_{j}^{\theta }=\widehat{X}_{j}^{\theta }-\langle \widehat{X}_{j}^{\theta }\rangle  \) and the frequency argument denotes a Fourier transform: \[
\widehat{X}_{j}^{\theta }(\omega )=\int \frac{dt}{\sqrt{2\pi }}e^{i\omega t}\widehat{X}_{j}^{\theta }(t)\, \, .\]

Since the P-representation is normally ordered, it automatically provides the normally-ordered moments: \begin{equation}
\langle :\widehat{x}_{j}^{\theta }\left( t\right) \widehat{x}_{j}^{\theta }\left( t\right) :\rangle =\langle {x}_{j}^{\theta }\left( t\right) {x}_{j}^{\theta }\left( t\right) \rangle _{P}\, \, .
\end{equation}
 Also, the +P spectral correlations correspond to the normally ordered, time-ordered operator correlations of the measured fields. We therefore define Fourier components of the normalized quadratures as: \begin{eqnarray}
x_{j}^{\theta }\left( t\right)  & = & \int \frac{d\omega }{\sqrt{2\pi }}e^{-i\omega t}\tilde{x}_{j}^{\theta }\left( \omega \right) \, \, ,
\end{eqnarray}

This leads to the following well-known result for the general squeezing spectrum, as measured in an external homodyne detection scheme: \begin{equation}
V_{j}^{\theta }(\omega )\delta (\omega +\omega ^{\prime })=1+2\gamma _{j}^{out}\left\langle \Delta \tilde{x}_{j}^{\theta }\left( \omega \right) \Delta \tilde{x}_{j}^{\theta }\left( \omega ^{\prime }\right) \right\rangle _{P}\, \, .
\end{equation}
 Note that vacuum (input) field terms do not contribute directly to this spectrum, as they have a vanishing normally-ordered spectrum, and are not correlated with the coherent amplitudes in the +P representation.

In the case of the Wigner representation, the correlations and moments are given with symmetric ordering. Thus, for example, \( \left\langle {\alpha }_{j}^{*}\left( t\right) {\alpha }_{j}\left( t\right) \right\rangle _{W}=\left\langle \left[ \widehat{a}_{j}\left( t\right) ,\widehat{a}_{j}^{\dagger }\left( t\right) \right] _{+}/2\right\rangle =1/2 \) in the vacuum state. The normally-ordered internal field moments are easily calculated, by using equal-time commutators to change the ordering from symmetric to normally-ordered: \begin{equation}
\left\langle :\widehat{x}_{j}^{\theta }\left( t\right) \widehat{x}_{j}^{\theta }\left( t\right) :\right\rangle =\left\langle {x}_{j}^{\theta }\left( t\right) {x}_{j}^{\theta }\left( t\right) \right\rangle _{W}-1\, \, .
\end{equation}

Similarly, the normally-ordered squeezing spectrum, as measured in an external homodyne detection scheme is: \begin{equation}
V_{j}^{\theta }(\omega )\delta (\omega +\omega ^{\prime })=\left\langle \tilde{X}_{j}^{\theta }\left( \omega \right) \tilde{X}_{j}^{\theta }\left( \omega ^{\prime }\right) \right\rangle _{W}\, \, .
\end{equation}
 It is essential here to include the vacuum field contributions from reflected input fields, as these are correlated with the internal Wigner amplitudes, and hence have a significant contribution to the spectrum. In fact, these input fields can be shown to correspond directly to the noise terms in the relevant Wigner equations, leading to the identification: \begin{equation}
\frac{dw_{j}}{dt}=\sqrt{2}{\Phi }^{in}_{j}(t)\, \, ,
\end{equation}
 where \( {\Phi }^{in}_{j}(t) \) is a c-number amplitude corresponding (in the Wigner representation) to the quantum vacuum input field.

The fundamental property of the Wigner function is that the ensemble average of any polynomial of the random variable \( a \) and \( a^{*} \) weighted by the Wigner density exactly corresponds to the Hilbert-space expectation of the corresponding symmetrized product of the annihilation and creation operators. Therefore, the truncated theory with a positive Wigner function can be viewed as equivalent to a hidden variable theory, since one can obtain quadrature fluctuation predictions by following an essentially classical prescription; in which even the noise terms have a classical interpretation as corresponding a form of zero-point fluctuation. This cannot be equivalent to quantum mechanics in general, but may provide similar results to quantum mechanics under some circumstances.

\section{Below-threshold perturbation theory}

Next we wish to rescale the equations. This has the merit of showing explicitly how a small noise expansion can permit us to use a type of perturbation theory whose zero-th order solution is the classical solution, rather than the Feynman approach where the zero-th order solution is the free-particle case. In order to show this systematically, a formal perturbation expansion in powers of \( g \) is now introduced, where the scaling parameter \( g \) is given by: \begin{equation}
g=1/\sqrt{4I_{c}\gamma _{1}}=1/\sqrt{2N_{c}\gamma _{r}}\, \, ,\end{equation}
 where \( N_{c}=2I_{c}/\gamma _{2} \) is the threshold pump photon number, and a dimensionless decay ratio, \( \gamma _{r}=\gamma _{2}/\gamma _{1} \) is introduced. An equivalent definition is:

\begin{equation}
g=\frac{\chi }{\sqrt{2\gamma _{1}\gamma _{2}}}\, \, ,
\end{equation}
 This clearly determines the ratio of nonlinear to linear rates of change. Next, we introduce a scaled time \( \tau =\gamma _{1}t \) and a dimensionless driving field \( \mu ={\mathcal{E}}/{\mathcal{E}}_{c}=\chi {\mathcal{E}}/(\gamma _{1}\gamma _{2}) \), so that the equations can be expressed in terms of the three dimensionless parameters \( g,\mu ,\gamma _{r} \). Finally, we expand the scaled coordinates in a power series in \( g \) , to give: \begin{eqnarray}
x_{1} & = & \sum _{n=0}^{\infty }g^{n-1}x_{1}^{\left( n\right) }\, \, ,\nonumber \\
y_{1} & = & \sum _{n=0}^{\infty }g^{n-1}y_{1}^{\left( n\right) }\, \, \nonumber \\
x_{2} & = & \frac{1}{\sqrt{2\gamma _{r}}}\sum _{n=0}^{\infty }g^{n-1}x_{2}^{\left( n\right) }\, \, ,\nonumber \\
y_{2} & = & \frac{1}{\sqrt{2\gamma _{r}}}\sum _{n=0}^{\infty }g^{n-1}y_{2}^{\left( n\right) }\, \, 
\end{eqnarray}

The expansion given here has the property that the zero-th order term corresponds to the large classical fields of order \( 1/g \), while the first order term corresponds to the quantum fluctuations of order \( 1 \), and the higher order terms correspond to nonlinear corrections to the quantum fluctuations, of order \( g \) and greater. For a given fundamental decay rate \( \gamma _{1} \), the expansion coefficient \( g^{2} \) is inversely proportional to the input photon flux required to obtain the threshold condition. Thus, the smaller \( g^{2} \) is, the larger the required input field.

\subsection{Matched power equations in positive P-representation}

Here we will first be interested in the analysis of the steady state moments. Subsequently we will calculate the spectral correlations of the solutions using Fourier transforms of the calculation done in the time domain. The equations for the quadrature variables in the positive P-representation are \begin{eqnarray}
dx_{1} & = & \left[ -\gamma _{1}x_{1}+\frac{\chi }{2}\left( x_{1}x_{2}+y_{1}y_{2}\right) \right] dt+\sqrt{\frac{\chi }{2}}\; \left[ \sqrt{x_{2}+iy_{2}}dw_{1}(t)+\sqrt{x_{2}^{\dagger }-iy_{2}^{\dagger }}dw_{2}(t)\right] \; \; ,\nonumber \\
dy_{1} & = & \left[ -\gamma _{1}y_{1}+\frac{\chi }{2}\left( x_{1}y_{2}-x_{2}y_{1}\right) \right] dt-i\sqrt{\frac{\chi }{2}}\; \left[ \sqrt{x_{2}+iy_{2}}dw_{1}(t)-\sqrt{x_{2}^{\dagger }-iy_{2}^{\dagger }}dw_{2}(t)\right] \; \; ,\nonumber \\
dx_{2} & = & \left[ -\gamma _{2}x_{2}-\frac{\chi }{4}\left( x_{1}^{2}-y_{1}^{2}\right) +2{\mathcal{E}}\right] dt\nonumber \\
dy_{2} & = & \left[ -\gamma _{2}y_{2}-\frac{\chi }{2}x_{1}y_{1}\right] dt\; \; .
\end{eqnarray}

The stochastic equations are now solved by the technique of matching powers of \( g \) in the corresponding time-evolution equations. This technique can be analyzed diagrammatically, and so can be termed the `stochastic diagram' method\cite{Stoch_diagram}. The zero-th order solution is: \begin{eqnarray}
dx_{1}^{\left( 0\right) } & = & \left[ -x_{1}^{\left( 0\right) }+\frac{1}{2}\left( x_{1}^{\left( 0\right) }x_{2}^{\left( 0\right) }+y_{1}^{\left( 0\right) }y_{2}^{\left( 0\right) }\right) \right] d\tau \, \, ,\nonumber \\
dy_{1}^{\left( 0\right) } & = & \left[ -y_{1}^{\left( 0\right) }+\frac{1}{2}\left( x_{1}^{\left( 0\right) }x_{2}^{\left( 0\right) }-x_{2}^{\left( 0\right) }y_{1}^{\left( 0\right) }\right) \right] d\tau \, \, ,\nonumber \\
dx_{2}^{\left( 0\right) } & = & -\gamma _{r}\left[ x_{2}^{\left( 0\right) }+\frac{1}{2}\left( x_{1}^{\left( 0\right) }x_{1}^{\left( 0\right) }-y_{1}^{\left( 0\right) }y_{1}^{\left( 0\right) }\right) -2\mu \right] d\tau \, \, ,\nonumber \\
dy_{2}^{\left( 0\right) } & = & -\gamma _{r}\left[ y_{2}^{\left( 0\right) }+x_{1}^{\left( 0\right) }y_{1}^{\left( 0\right) }\right] d\tau \, \, .
\end{eqnarray}

These equations are the classical nonlinear equations for the cavity, expressed in terms of the quadrature amplitudes of dimensionless scaled fields. The steady-state solution below threshold is well-known, and is given by:

\begin{equation}
x_{1}^{\left( 0\right) }=y_{1}^{\left( 0\right) }=y_{2}^{\left( 0\right) }=0;\, \, \, x_{2}^{\left( 0\right) }=2\mu \, \, .
\end{equation}
 With no loss of generality, we can set all odd orders of \( x_{2}^{\left( n\right) },y_{2}^{\left( n\right) } \), and all even orders of \( x_{1}^{\left( n\right) },y_{1}^{\left( n\right) } \) to zero, since one can set these to zero initially, and these orders do not change in time. To first order, the equations are given by: \begin{eqnarray}
dx_{1}^{\left( 1\right) } & = & -\left( 1-\mu \right) x_{1}^{\left( 1\right) }d\tau +\sqrt{2\mu }dw_{x}(\tau )\, \, ,\nonumber \\
dy_{1}^{\left( 1\right) } & = & -\left( 1+\mu \right) y_{1}^{\left( 1\right) }d\tau -i\sqrt{2\mu }dw_{y}(\tau )\, \, ,\label{2} 
\end{eqnarray}
 where, \( dw_{x(y)}(\tau )=\left( dw_{1}(\tau )\pm dw_{2}(\tau )\right) /\sqrt{2} \). These equations are the ones that are normally used to predict squeezing. They are non-classical, but correspond to a very simple form of linear, non-classical fluctuation which has a Gaussian quasi-probability distribution. In other words, if no higher-order terms existed, the result would be an ideal squeezed state in the sub-harmonic, together with an ideal coherent state in the pump.

Of more interest to the present paper, is the behaviour to the next order. This is the first order where nonlinear corrections to ideal squeezed-state behaviour will occur. We find the following:

\noindent \begin{eqnarray}
dx_{2}^{\left( 2\right) } & = & -\gamma _{r}\left[ x_{2}^{\left( 2\right) }+\frac{1}{2}\left( x_{1}^{\left( 1\right) }x_{1}^{\left( 1\right) }-y_{1}^{\left( 1\right) }y_{1}^{\left( 1\right) }\right) \right] d\tau \, \, ,\nonumber \\
dy_{2}^{\left( 2\right) } & = & -\gamma _{r}\left[ y_{2}^{\left( 2\right) }+x_{1}^{\left( 1\right) }y_{1}^{\left( 1\right) }\right] d\tau \, \, .\label{8} 
\end{eqnarray}
 While we do not wish to include any effects beyond the first nonlinear corrections, it is not possible to consistently neglect the third-order in perturbation theory. This is because the first non-trivial correlations arise in terms like \( \left\langle \left[ x^{\left( 2\right) }\right] ^{2}\right\rangle , \) which have the same formal order as terms of the type \( \left\langle x^{\left( 3\right) }x^{\left( 1\right) }\right\rangle . \) Therefore, to obtain a consistent expansion for the correlations that are of interest, we must compute the third-order terms as well. These satisfy the following equations: \begin{eqnarray}
dx_{1}^{\left( 3\right) } & = & \left[ -\left( 1-\mu \right) x_{1}^{\left( 3\right) }+\frac{1}{2}\left( x_{1}^{\left( 1\right) }x_{2}^{\left( 2\right) }+y_{1}^{\left( 1\right) }y_{2}^{\left( 2\right) }\right) \right] d\tau +\nonumber \\
 & + & \frac{1}{2\sqrt{2\mu }}\left[ x_{2}^{\left( 2\right) }dw_{x}(\tau )+iy_{2}^{\left( 2\right) }dw_{y}(\tau )\right] \, \, ,\nonumber \\
dy_{1}^{\left( 3\right) } & = & \left[ -\left( 1+\mu \right) y_{1}^{\left( 3\right) }+\frac{1}{2}\left( x_{1}^{\left( 1\right) }y_{2}^{\left( 2\right) }-x_{2}^{\left( 2\right) }y_{1}^{\left( 1\right) }\right) \right] d\tau \nonumber \\
 & + & \frac{1}{2\sqrt{2\mu }}\left[ y_{2}^{\left( 2\right) }dw_{x}(\tau )-ix_{2}^{\left( 2\right) }dw_{y}(\tau )\right] \, \, .\label{10} 
\end{eqnarray}
 The equations of this order have a non-trivial noise term, which depends on the second order pump quadrature solution.

\subsubsection{Operator moments}

We now wish to calculate the operator moments. To proceed further, we use It\^{o} calculus to derive stochastic equations for quantities of interest, which in the present calculation are \( y_{1}^{\left( 1\right) }y_{1}^{\left( 1\right) } \) and \( y_{1}^{\left( 1\right) }y_{1}^{\left( 3\right) } \). These equations contain quantities involving variables lower down in the hierarchy, as well as terms generated from the noise correlations. Finally, we compute the the steady state averages of the quantities of interest, so that the noise terms vanish. In the present case, this yields, \begin{eqnarray}
\left\langle x_{2}^{\left( 2\right) }\right\rangle  & = & -\frac{\mu }{1-\mu ^{2}}\, \, ,\nonumber \\
\left\langle y_{1}^{\left( 1\right) }y_{1}^{\left( 1\right) }\right\rangle  & = & -\frac{\mu }{1+\mu }\, \, ,\nonumber \\
\left\langle x_{1}^{\left( 1\right) }x_{1}^{\left( 1\right) }\right\rangle  & = & \frac{\mu }{1-\mu },\, \, ,\nonumber \\
\left\langle y_{1}^{\left( 1\right) }y_{1}^{\left( 3\right) }\right\rangle  & = & \frac{\mu }{4\left( 1+\mu \right) \left( 1-\mu ^{2}\right) }\left[ \frac{\mu \gamma _{r}}{\gamma _{r}+2}+\frac{\gamma _{r}\left( 1-\mu +\mu ^{2}\right) +2\left( 1+\mu \right) }{\left( 1+\mu \right) \left( \gamma _{r}+2\left( 1+\mu \right) \right) }\right] \, \, ,\nonumber \\
\left\langle x_{1}^{\left( 1\right) }y_{1}^{\left( 1\right) }y_{2}^{\left( 2\right) }\right\rangle  & = & \frac{\gamma _{r}}{\left( \gamma _{r}+2\right) }\left( \frac{\mu ^{2}}{1-\mu ^{2}}\right) \, \, .
\end{eqnarray}

The first quantity above is related to the depletion of the pump that supplies the energy for the sub-harmonic mode. The follow two quantities are the squeezed and enhanced quadratures normally obtained in the linearized theory, while the fourth one is the first correction to the linearized calculation. The last one is the steady state triple quadrature correlation. This quantity has been suggested previously as a way to test quantum mechanics against a local hidden variable theory~\cite{Triple}.

The steady state intra-cavity squeezed quadrature fluctuations are obtained as: \begin{eqnarray}
{\left\langle {\widehat{y}_{1}}^{2}\right\rangle }_{ss} & = & 1+\left\langle :{\widehat{y}_{1}}^{2}:\right\rangle \nonumber \\
 & = & \frac{1}{\left( 1+\mu \right) }+\frac{g^{2}\mu }{2\left( 1+\mu \right) ^{2}\left( 1-\mu \right) }\times \nonumber \\
 & \times  & \left[ \frac{\mu \gamma _{r}}{\gamma _{r}+2}+\frac{\gamma _{r}\left( 1-\mu +\mu ^{2}\right) +2\left( 1+\mu \right) }{\left( 1+\mu \right) \left( \gamma _{r}+2\left( 1+\mu \right) \right) }\right] \, \, .
\end{eqnarray}

\begin{figure}
{\centering \resizebox*{10cm}{10cm}{\includegraphics{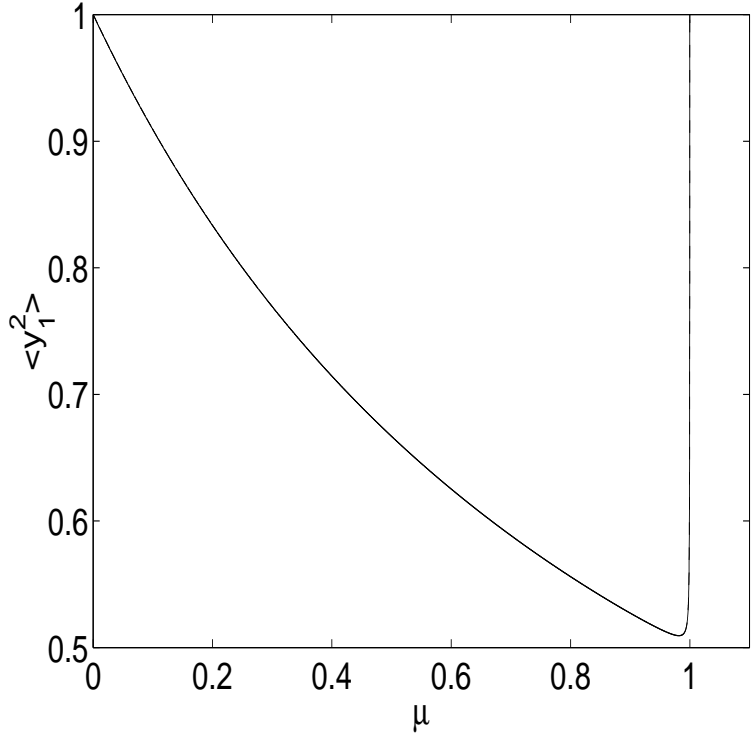}} \par}

\caption{Squeezing moment \protect\( \left\langle {\widehat{y}_{1}}^{2}\right\rangle \protect \) versus driving field \protect\( \mu \protect \), with \protect\( g^{2}=0.001,\protect \) \protect\( \gamma _{r}=0.5\protect \) .}

\label{Sqmom}
\end{figure}

The intra-cavity squeezing quadrature near threshold is not perfectly squeezed, as the nonlinear correction is divergent near this point. This is shown in Fig. (\ref{Sqmom}). It is clear that the nonlinear corrections to the overall moment scale as \( g^{2}/(1-\mu ) \) , and hence only give large corrections extremely close to threshold, with \( \mu \approx 1-g^{2} \) .

Considerations related to optimal squeezing will be treated later, in the frequency domain.

\subsection{Matched power equations in semiclassical theory}

\noindent We can scale the quadratures variables in in semiclassical theory in the same way as before. Firstly, the equations for the quadratures are:

\noindent \begin{eqnarray}
dx_{1} & = & \left[ -\gamma _{1}x_{1}+\frac{\chi }{2}\left( x_{1}x_{2}+y_{1}y_{2}\right) \right] dt+\sqrt{\gamma _{1}}\; \left[ dw_{1}(t)+dw_{1}^{*}(t)\right] \; \; ,\nonumber \\
dy_{1} & = & \left[ -\gamma _{1}y_{1}+\frac{\chi }{2}\left( x_{1}y_{2}-x_{2}y_{1}\right) \right] dt-i\sqrt{\gamma _{1}}\left[ dw_{1}(t)+dw_{1}^{*}(t)\right] \; \; ,\nonumber \\
dx_{2} & = & \left[ -\gamma _{2}x_{2}-\frac{\chi }{4}\left( x_{1}^{2}-y_{1}^{2}\right) +2{\mathcal{E}}\right] dt+\sqrt{\gamma _{2}}\; \left[ dw_{2}(t)+dw_{2}^{*}(t)\right] \; \; ,\nonumber \\
dy_{2} & = & \left[ -\gamma _{2}y_{2}-\frac{\chi }{2}x_{1}y_{1}\right] dt-i\sqrt{\gamma _{2}}\; \left[ dw_{2}(t)-dw_{2}^{*}(t)\right] \; \; .
\end{eqnarray}

In the new scaled time, the correlation function of the noise terms is: \begin{equation}
\langle \xi _{k}(t)\xi _{l}^{*}(t^{\prime })\rangle =\langle \xi _{k}\left( \tau /\gamma _{1}\right) \xi _{l}^{*}\left( \tau ^{\prime }/\gamma _{1}\right) \rangle =\gamma _{1}\delta _{kl}\delta (\tau -\tau ^{\prime })=\gamma _{1}\langle \xi _{k}(\tau )\xi _{l}^{*}(\tau ^{\prime })\rangle \, \, ,
\end{equation}
 where we have written the Wiener increment as \( dw(t)=\xi (t)dt \). Next, we redefine the white noise that drives the stochastic semiclassical equations as \begin{eqnarray}
dw_{x1(2)}(\tau ) & = & \frac{\left[ dw_{1(2)}(\tau )+dw_{1(2)}^{*}(\tau )\right] }{\sqrt{2}}\, \, ,\nonumber \\
dw_{y1(2)}(\tau ) & = & \frac{\left[ dw_{1(2)}(\tau )-dw_{1(2)}^{*}(\tau )\right] }{i\sqrt{2}}\, \, .
\end{eqnarray}

The dimensionless driving field \( \mu  \) is introduced as before, and the Wiener increments \( dw_{i}(\tau ) \) have the same properties as defined in (\ref{a8}), except for changing \( t \) to the dimensionless scaled time \( \tau  \). Next, we use the same technique of matching the powers of g in the corresponding time-evolution equations. The zero-th order equations are: \begin{eqnarray}
dx_{1}^{(0)} & = & \left[ -x_{1}^{(0)}+\frac{1}{2}\left( x_{1}^{(0)}x_{2}^{(0)}+y_{1}^{(0)}y_{2}^{(0)}\right) \right] d\tau \, \, ,\nonumber \\
dy_{1}^{(0)} & = & \left[ -y_{1}^{(0)}+\frac{1}{2}\left( x_{1}^{(0)}y_{2}^{(0)}-x_{2}^{(0)}y_{1}^{(0)}\right) \right] d\tau \, \, ,\nonumber \\
dx_{2}^{(0)} & = & -\gamma _{r}\left[ x_{2}^{(0)}+\frac{1}{2}\left( {x_{1}^{(0)}}{x_{1}^{(0)}}-{y_{1}^{(0)}}{y_{1}^{(0)}}\right) -2\mu \right] d\tau \, \, ,\nonumber \\
dy_{2}^{(0)} & = & -\gamma _{r}\left[ y_{2}^{(0)}+x_{1}^{(0)}y_{1}^{(0)}\right] d\tau \, \, .
\end{eqnarray}
 As in the positive-P case, the steady state solution below threshold is given by: \begin{equation}
x_{1}^{(0)}=y_{1}^{(0)}=y_{2}^{(0)}=0\; \; \; \; ;\; \; \; \; x_{2}^{(0)}=2\mu \, \, .
\end{equation}
 To first order, the equations are given by: \begin{eqnarray}
dx_{1}^{(1)} & = & -\left( 1-\mu \right) x_{1}^{(1)}d\tau +\sqrt{2}dw_{x1}(\tau )\, \, ,\nonumber \\
dy_{1}^{(1)} & = & -\left( 1+\mu \right) y_{1}^{(1)}d\tau +\sqrt{2}dw_{y1}(\tau )\, \, ,\nonumber \\
dx_{2}^{(1)} & = & -\gamma _{r}x_{2}^{(1)}d\tau +2\gamma _{r}dw_{x2}(\tau )\, \, ,\nonumber \\
dy_{2}^{(1)} & = & -\gamma _{r}y_{2}^{(1)}d\tau +2\gamma _{r}dw_{y2}(\tau )\, \, .
\end{eqnarray}

While the zero-th order equations are essentially classical, in this first order set the noise appears as a quantum effect. This is still a linear approximation, as all nonlinear corrections come from the next orders.

The second order equations are: \begin{eqnarray}
dx_{1}^{(2)} & = & \left[ -\left( 1-\mu \right) x_{1}^{(2)}+\frac{1}{2}\left( x_{1}^{(1)}x_{2}^{(1)}+y_{1}^{(1)}y_{2}^{(1)}\right) \right] d\tau \, \, ,\nonumber \\
dy_{1}^{(2)} & = & \left[ -\left( 1+\mu \right) y_{1}^{(2)}+\frac{1}{2}\left( x_{1}^{(1)}y_{2}^{(1)}-x_{2}^{(1)}y_{1}^{(1)}\right) \right] d\tau \, \, ,\nonumber \\
dx_{2}^{(2)} & = & -\gamma _{r}\left[ x_{2}^{(2)}+\frac{1}{2}\left( {x_{1}^{(1)}}{x_{1}^{(1)}}-{y_{1}^{(1)}}{y_{1}^{(1)}}\right) \right] d\tau \, \, ,\nonumber \\
dy_{2}^{(2)} & = & -\gamma _{r}\left[ y_{2}^{(2)}+x_{1}^{(1)}y_{1}^{(1)}\right] d\tau \, \, .
\end{eqnarray}

We need to go beyond this order in perturbation theory to compute the first nonlinear corrections. The third order equations are: \begin{eqnarray}
dx_{1}^{(3)} & = & \left[ -\left( 1-\mu \right) x_{1}^{(3)}+\frac{1}{2}\left( x_{1}^{(1)}x_{2}^{(2)}+x_{1}^{(2)}x_{2}^{(1)}+y_{1}^{(1)}y_{2}^{(2)}+y_{1}^{(2)}y_{2}^{(1)}\right) \right] d\tau \, \, ,\nonumber \\
dy_{1}^{(3)} & = & \left[ -\left( 1+\mu \right) y_{1}^{(3)}+\frac{1}{2}\left( x_{1}^{(1)}y_{2}^{(2)}+x_{1}^{(2)}y_{2}^{(1)}-x_{2}^{(1)}y_{1}^{(2)}-x_{2}^{(2)}y_{1}^{(1)}\right) \right] d\tau \, \, ,\nonumber \\
dx_{2}^{(3)} & = & -\gamma _{r}\left[ x_{2}^{(3)}+\left( {x_{1}^{(1)}}{x_{1}^{(2)}}-{y_{1}^{(1)}}{y_{1}^{(2)}}\right) \right] d\tau \, \, ,\nonumber \\
\frac{dy_{2}^{(3)}}{d\tau } & = & -\gamma _{r}\left[ y_{2}^{(3)}+x_{1}^{(1)}y_{1}^{(2)}+x_{1}^{(2)}y_{1}^{(1)}\right] d\tau \, \, .
\end{eqnarray}

\subsubsection{Operator moments}

The steady state averages of the quantities of interest can now be using the truncated Wigner distribution, therefore obtaining the symmetrically ordered correlation functions: \begin{eqnarray}
\langle x_{2}^{(2)}\rangle  & = & -\frac{\mu }{1-\mu ^{2}}\, \, ,\nonumber \\
\langle y_{1}^{(1)}y_{1}^{(1)}\rangle  & = & \frac{1}{1+\mu }\, \, ,\nonumber \\
\langle x_{1}^{(1)}x_{1}^{(1)}\rangle  & = & \frac{1}{1-\mu }\, \, ,\nonumber \\
\langle y_{1}^{(2)}y_{1}^{(2)}\rangle  & = & \frac{1}{2}\left( \frac{\gamma _{r}}{\gamma _{r}+2}\right) \frac{1}{(1-\mu ^{2})}+\frac{\gamma _{r}}{2(1+\mu )^{2}[\gamma _{r}+2(1+\mu )]}\, \, ,\nonumber \\
\langle y_{1}^{(1)}y_{1}^{(3)}\rangle  & = & \frac{\mu }{4(1+\mu )(1-\mu ^{2})}\left[ \frac{-\gamma _{r}}{\gamma _{r}+2}+\frac{\gamma _{r}(2-\mu )+2(1+\mu )}{(1+\mu )[\gamma _{r}+2(1+\mu )]}\right] \, \, ,\nonumber \\
\langle x_{1}^{(1)}y_{1}^{(1)}y_{2}^{(2)}\rangle  & = & -\left( \frac{\gamma _{r}}{\gamma _{r}+2}\right) \left( \frac{1}{1-\mu ^{2}}\right) \, \, ,\nonumber \\
\langle x_{1}^{(1)}y_{1}^{(1)}y_{2}^{(2)}\rangle +\langle x_{1}^{(2)}y_{1}^{(1)}y_{2}^{(1)}\rangle  & = & \left( \frac{2\gamma _{r}}{\gamma _{r}+2}\right) \left( \frac{1}{1-\mu ^{2}}\right) \, \, .
\end{eqnarray}

The main difference in these calculation compared with the positive-P result, appears in the nonlinear correction for the sub-harmonic squeezed quadrature. Up to second order in \( g \) we have \begin{eqnarray}
\langle \widehat{y}_{1}^{2}\rangle  & = & \frac{1}{g^{2}}\left[ {g^{2}}\langle y_{1}^{(1)}y_{1}^{(1)}\rangle +{g^{4}}\langle y_{1}^{(2)}y_{1}^{(2)}\rangle +2{g^{4}}\langle y_{1}^{(1)}y_{1}^{(3)}\rangle \right] \nonumber \\
 & = & \frac{1}{1+\mu }+ \nonumber \\
 & + & \frac{g^{2}}{2(1+\mu )(1-\mu ^{2})}\left[ \frac{\gamma _{r}}{\gamma _{r}+2}+\frac{\gamma _{r}(1+2\mu -2\mu ^{2})+2\mu (1+\mu )}{(1+\mu )[\gamma _{r}+2(1+\mu )]}\right] \, \, .
\end{eqnarray}

The similarities and disagreement between this result and the positive-P expression for the same quantity deserve further comments given in the conclusion section. In particular, we notice that while the linear term agrees, the nonlinear term is not in agreement well below threshold. 

\begin{figure}
{\centering \resizebox*{10cm}{10cm}{\includegraphics{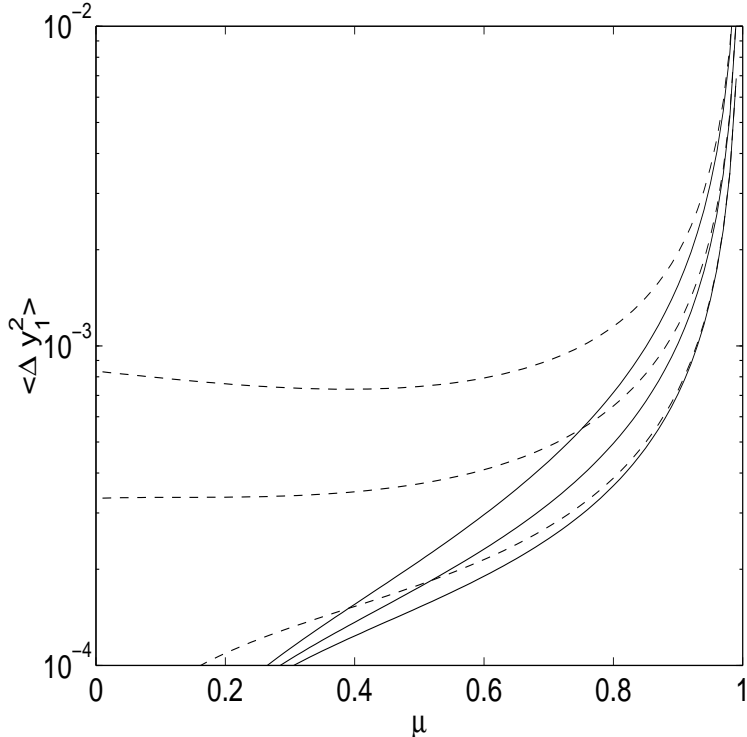}} \par}

\caption{Nonlinear correction to the squeezing moment \protect\( \left\langle \Delta {\widehat{y}_{1}}^{2}\right\rangle \protect \) versus driving field \protect\( \mu \protect \), with \protect\( g^{2}=0.001,\protect \) \protect\( \gamma _{r}=0.1,1,10 \protect \) . Solid line corresponds to the positive-P representation, dashed line to the Wigner representation. Best squeezing occurs with smallest value of \protect\( \gamma _{r}.\protect \)}

\label{NLSqmom}
\end{figure}

This comparison is shown in Fig(\ref{NLSqmom}), which compares the nonlinear parts of the moment in the two representations.

Just below threshold both theories give nonlinear corrections which are essentially identical. There is also good agreement in the limit of \( \gamma _{r}\rightarrow 0 \), where \( \gamma _{2}\ll \gamma _{1} \); but for \( \gamma _{r}>0 \) and driving fields below threshold, there is substantial disagreement in the nonlinear corrections to the squeezing between the two representations. This can be attributed to the neglect of third-order quantum correlations in the truncated Wigner representation, which results in the appearance of nonlinear squeezing effects even in the limit of zero driving field. Such effects are due to the semiclassical vacuum inputs, which do not appear in the positive P-representation.

\section{Spectral Correlations}

Next, we proceed to analyze the problem in the frequency space by taking the Fourier decomposition of the fields in order to understand the role of the first nonlinear correction in the squeezing spectrum. It is important to stress that most of the measurements performed are done in Fourier space.

The nonlinear corrections to the spectrum have a strikingly different behaviour to the case of the squeezing moments. The reason for this is that the nonlinear corrections are due to low-frequency, narrow-band critical fluctuations. These have a very small effect on the moments, which correspond to an integral of the spectrum over all frequencies, unless extremely close to threshold. However, they can have a very large and disruptive effect on the very important zero frequency component of the squeezing spectrum, where the quantum noise is at its lowest level.

\subsection{Positive P-representation}

The spectrum can be calculated directly from the Fourier transform of the stochastic equations. We also represent the white noise that drives the stochastic equations by its Fourier transform \( \xi _{x,y}\left( \Omega \right)  \), where the spectral moments of the stochastic processes are: \begin{eqnarray}
\left\langle \xi _{a}\left( \Omega \right) \right\rangle  & = & 0\, \, ,\nonumber \\
\left\langle \xi _{a}\left( \Omega \right) \xi _{b}\left( \Omega ^{\prime }\right) \right\rangle  & = & \delta _{ab}\delta \left( \Omega +\Omega ^{\prime }\right) \, \, .
\end{eqnarray}
 It is also useful to introduce a standard convolution notation, where:

\begin{equation}
[A\star B](\Omega )=\int \frac{d\Omega ^{\prime }}{\sqrt{2\pi }}A(\Omega ^{\prime })B(\Omega -\Omega ^{\prime })\, \, .\end{equation}
 The stochastic equations may now be rewritten in the frequency domain as -

\begin{itemize}
\item First order: \begin{eqnarray}
{\tilde{x}}_{1}^{\left( 1\right) }\left( \Omega \right)  & = & \frac{\sqrt{2\mu }\xi _{x}\left( \Omega \right) }{\left( i\Omega +1-\mu \right) }\, \, ,\nonumber \\
{\tilde{y}}_{1}^{\left( 1\right) }\left( \Omega \right)  & = & -\frac{i\sqrt{2\mu }\xi _{y}\left( \Omega \right) }{\left( i\Omega +1+\mu \right) }\, \, .\label{14} 
\end{eqnarray}

\item Second order: \begin{eqnarray}
{\tilde{x}}_{2}^{\left( 2\right) }\left( \Omega \right)  & = & -\frac{\gamma _{r}\left[ \tilde{x}_{1}^{\left( 1\right) }\star \tilde{x}_{1}^{\left( 1\right) }-\tilde{y}_{1}^{\left( 1\right) }\star \tilde{y}_{1}^{\left( 1\right) }\right] \left( \Omega \right) }{2\left( i\Omega +\gamma _{r}\right) }\, \, ,\nonumber \\
{\tilde{y}}_{2}^{\left( 2\right) }\left( \Omega \right)  & = & -\frac{\gamma _{r}\left[ \tilde{x}_{1}^{\left( 1\right) }\star \tilde{y}_{1}^{\left( 1\right) }\right] \left( \Omega \right) }{\left( i\Omega +\gamma _{r}\right) }\, \, .\label{19} 
\end{eqnarray}

\item Third order: 
\end{itemize}
\begin{eqnarray}
{\tilde{x}}_{1}^{\left( 3\right) }\left( \Omega \right)  & = & \frac{\left[ \tilde{x}_{2}^{\left( 2\right) }\star \left( \tilde{x}_{1}^{\left( 1\right) }+\xi _{x}/\sqrt{2\mu }\right) +\tilde{y}_{2}^{\left( 2\right) }\star \left( \tilde{y}_{1}^{\left( 1\right) }+i\xi _{y}/\sqrt{2\mu }\right) \right] \left( \Omega \right) }{2\left( i\Omega +1-\mu \right) }\, \, ,\nonumber \\
{\tilde{y}}_{1}^{\left( 3\right) }\left( \Omega \right)  & = & \frac{\left[ \tilde{y}_{2}^{\left( 2\right) }\star \left( \tilde{x}_{1}^{\left( 1\right) }+\xi _{x}/\sqrt{2\mu }\right) -\tilde{x}_{2}^{\left( 2\right) }\star \left( \tilde{y}_{1}^{\left( 1\right) }+i\xi _{y}/\sqrt{2\mu }\right) \right] \left( \Omega \right) }{2\left( i\Omega +1+\mu \right) }\, \, .
\end{eqnarray}

\subsubsection{Squeezing correlation spectrum}

We now calculate the spectrum of the squeezed field \( y_{1} \), which is given by \( \left\langle \tilde{y}_{1}\left( \Omega _{1}\right) \tilde{y}_{1}\left( \Omega _{2}\right) \right\rangle  \). Thus, we obtain

\begin{eqnarray}
\left\langle \tilde{y}_{1}\left( \Omega _{1}\right) \tilde{y}_{1}\left( \Omega _{2}\right) \right\rangle  & = & \left\langle \tilde{y}_{1}^{\left( 1\right) }\left( \Omega _{1}\right) \tilde{y}_{1}^{\left( 1\right) }\left( \Omega _{2}\right) \right\rangle +\nonumber \\
 & + & g^{2}\left\langle \tilde{y}_{1}^{\left( 1\right) }\left( \Omega _{2}\right) \tilde{y}_{1}^{\left( 3\right) }\left( \Omega _{1}\right) +[\Omega _{1}\leftrightarrow \Omega _{2}]\right\rangle +\cdots 
\end{eqnarray}
 The contribution from the first order perturbation theory is the usual linearized squeezing result, given in this case by: \begin{equation}
\left\langle \tilde{y}_{1}^{\left( 1\right) }\left( \Omega _{1}\right) \tilde{y}_{1}^{\left( 1\right) }\left( \Omega _{2}\right) \right\rangle =-\frac{2\mu \delta \left( \Omega _{1}+\Omega _{2}\right) }{\left[ \Omega _{1}^{2}+\left( 1+\mu \right) ^{2}\right] }\, \, .
\end{equation}
 Similarly, the complementary (unsqueezed) spectrum is: \begin{equation}
\left\langle \widetilde{x}_{1}^{\left( 1\right) }\left( \Omega _{1}\right) \widetilde{x}_{1}^{\left( 1\right) }\left( \Omega _{2}\right) \right\rangle =\frac{2\mu \delta \left( \Omega _{1}+\Omega _{2}\right) }{\left[ \Omega _{1}^{2}+\left( 1-\mu \right) ^{2}\right] }\, \, .
\end{equation}
 Also, we can obtain the next order contribution to the squeezing, by calculating \( \left\langle \tilde{y}_{1}^{\left( 3\right) }\left( \Omega _{1}\right) \tilde{y}_{1}^{\left( 1\right) }\left( \Omega _{2}\right) \right\rangle \, \,  \). To check the results, we can compare with the moment calculations, since: \begin{eqnarray}
\left\langle {y}_{1}^{\left( 1\right) }\left( t\right) {y}_{1}^{\left( 3\right) }\left( t\right) \right\rangle _{ss}=\int \frac{d\Omega _{1}}{\sqrt{2\pi }}\int \frac{d\Omega _{2}}{\sqrt{2\pi }}\left\langle \tilde{y}_{1}^{\left( 1\right) }\left( \Omega _{1}\right) \tilde{y}_{1}^{\left( 3\right) }\left( \Omega _{2}\right) \right\rangle \, \, .
\end{eqnarray}
 Using these results, we find that the internal spectrum of the squeezed quadrature, to this order, is given by \begin{equation}
\left\langle \tilde{y}_{1}\left( \Omega _{1}\right) \tilde{y}_{1}\left( \Omega _{2}\right) \right\rangle =\delta (\Omega _{1}+\Omega _{2})S(\Omega _{1})\, \, ,
\end{equation}
 and the squeezing spectrum is calculated to be: \begin{eqnarray}
S(\Omega ) & = & \frac{-2\mu }{\Omega ^{2}+(1+\mu )^{2}}+\frac{2g^{2}\mu ^{2}\gamma _{r}}{[\Omega ^{2}+(1+\mu )^{2}]^{2}}\nonumber \\
 & \times  & \left[ \frac{(\Omega ^{2}+1-\mu ^{2})}{2\mu \gamma _{r}(1-\mu ^{2})}+\frac{(1-\mu +\gamma _{r})(1+\mu )-\Omega ^{2}}{(1-\mu )[\Omega ^{2}+(1-\mu +\gamma _{r})^{2}]}\right. \nonumber \\
 &  & \left. -\frac{(1+\mu +\gamma _{r})(1+\mu )-\Omega ^{2}}{(1+\mu )[\Omega ^{2}+(1+\mu +\gamma _{r})^{2}]}\right] \, \, .
\end{eqnarray}

\begin{figure}
{\centering \resizebox*{10cm}{10cm}{\includegraphics{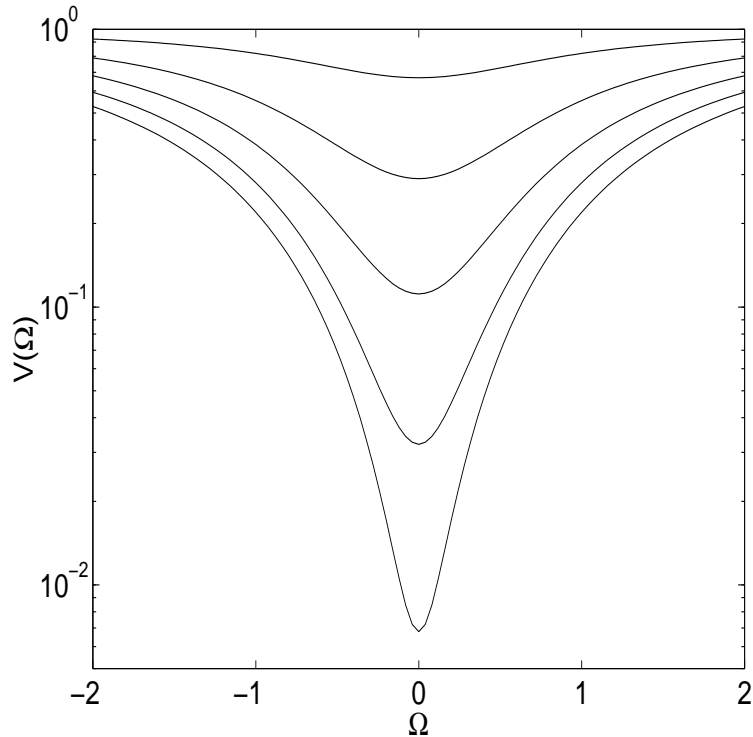}} \par}

\caption{Total OPO squeezing spectrum with \protect\( g^{2}=0.001\protect \) ,\protect\( \gamma _{r}=0.5.\protect \) The \protect\( \mu \protect \) values plotted are \protect\( \mu =0.1,0.3,0.5,0.7,0.9\protect \); larger values of  \( \mu  \)
give the most squeezing (lowest spectral variance). \label{TOTALSPEC}}
\end{figure}

The corresponding external squeezing spectrum is then:

\begin{eqnarray}
V(\Omega ) & = & 1-\frac{4\mu }{\Omega ^{2}+(1+\mu )^{2}}+\frac{4g^{2}\mu ^{2}\gamma _{r}}{[\Omega ^{2}+(1+\mu )^{2}]^{2}}\nonumber \\
 & \times  & \left[ \frac{(\Omega ^{2}+1-\mu ^{2})}{2\mu \gamma _{r}(1-\mu ^{2})}+\frac{(1-\mu +\gamma _{r})(1+\mu )-\Omega ^{2}}{(1-\mu )[\Omega ^{2}+(1-\mu +\gamma _{r})^{2}]}\right. \nonumber \\
 &  & \left. -\frac{(1+\mu +\gamma _{r})(1+\mu )-\Omega ^{2}}{(1+\mu )[\Omega ^{2}+(1+\mu +\gamma _{r})^{2}]}\right] \, \, .
\end{eqnarray}
 This equation gives the complete linear and nonlinear squeezing spectrum, including all the nonlinear correction terms that contribute to order \( g^{2} \) or \( 1/N \). An illustration of the behaviour of the total spectrum is given in Fig (\ref{TOTALSPEC}).

\begin{figure}
{\centering \resizebox*{10cm}{10cm}{\includegraphics{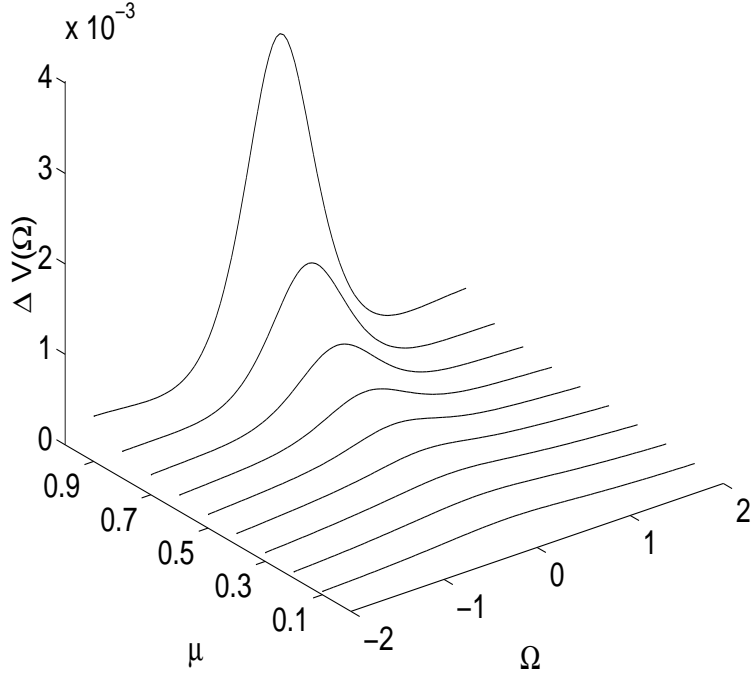}} \par}

\caption{Nonlinear OPO squeezing spectrum with \protect\( g^{2}=0.001\protect \) ,\protect\( \gamma _{r}=0.5.\protect \) The maximum \protect\( \mu \protect \) value plotted is \protect\( \mu =0.95\protect \) . \label{NLSPEC}}
\end{figure}

~Fig (\ref{NLSPEC}) shows how the nonlinear contribution changes with driving field, giving just the portion of the spectrum proportional to \( g^{2} \) .

\subsubsection{Triple Spectral Correlations}

Next, we can calculate the triple spectral correlations, giving as in the moment calculations:

\begin{eqnarray}
\left\langle \tilde{x}_{1}\left( \Omega _{1}\right) \tilde{y}_{1}\left( \Omega _{2}\right) \tilde{y}_{2}\left( \Omega _{3}\right) \right\rangle  & = & g \left\langle \tilde{x}_{1}^{\left( 1\right) }\left( \Omega _{1}\right) \tilde{y}_{1}^{\left( 1\right) }\left( \Omega _{2}\right) \tilde{y}_{2}^{\left( 2\right) }\left( \Omega _{3}\right) \right\rangle \, \, .
\end{eqnarray}

Solving for \( \tilde{y}_{2}^{\left( 2\right) } \) , we have \begin{eqnarray}
\left\langle \tilde{x}_{1}^{\left( 1\right) }\left( \Omega _{1}\right) \tilde{y}_{1}^{\left( 1\right) }\left( \Omega _{2}\right) \tilde{y}_{2}^{\left( 2\right) }\left( \Omega _{3}\right) \right\rangle  & = & -\frac{\gamma _{r}\left\langle \tilde{x}_{1}^{\left( 1\right) }\left( \Omega _{1}\right) \tilde{y}_{1}^{\left( 1\right) }\left( \Omega _{2}\right) [\tilde{x}_{1}^{\left( 1\right) }\star \tilde{y}_{1}^{\left( 1\right) }](\Omega _{3})\right\rangle }{\left( i\Omega _{3}+\gamma _{r}\right) }\, \, .
\end{eqnarray}

Substituting from the first order spectrum, the final result to this order is obtained to be: \begin{equation}
\left\langle \tilde{x}_{1}^{\left( 1\right) }\left( \Omega _{1}\right) \tilde{y}_{1}^{\left( 1\right) }\left( \Omega _{2}\right) \tilde{y}_{2}^{\left( 2\right) }\left( \Omega _{3}\right) \right\rangle =\frac{4\mu ^{2}\gamma _{r}/\sqrt{2\pi }\; \delta \left( \Omega _{1}+\Omega _{2}+\Omega _{3}\right) }{\left( i\Omega _{3}+\gamma _{r}\right) \left[ \Omega _{1}^{2}+\left( 1-\mu \right) ^{2}\right] \left[ \Omega _{2}^{2}+\left( 1+\mu \right) ^{2}\right] }\, \, .
\end{equation}

To check this result, we can evaluate moments: \begin{eqnarray}
\left\langle {x}_{1}^{\left( 1\right) }\left( t\right) {y}_{1}^{\left( 1\right) }\left( t\right) {y}_{2}^{\left( 2\right) }\left( t\right) \right\rangle _{ss} & = & \int \frac{d\Omega _{1}}{\sqrt{2\pi }}\int \frac{d\Omega _{2}}{\sqrt{2\pi }}\int \frac{d\Omega _{3}}{\sqrt{2\pi }}e^{i\left( \Omega _{1}+\Omega _{2}+\Omega _{3}\right) }\times \nonumber \\
\times \left\langle \tilde{x}_{1}^{\left( 1\right) }\left( \Omega _{1}\right) \tilde{y}_{1}^{\left( 1\right) }\left( \Omega _{2}\right) \tilde{y}_{2}^{\left( 2\right) }\left( \Omega _{3}\right) \right\rangle \, \, .
\end{eqnarray}
 On integrating, we obtain the same result as in our moment calculation, given above.

\subsection{Semiclassical theory}

We will now compare these results with the corresponding results calculated in the semiclassical theory. Some differences between them could be an interesting test comparing quantum mechanical predictions with a hidden variable theory.

Again, the spectral correlations are calculated from the Fourier transform of the stochastic equations. In the frequency domain, the equations are written as

\begin{itemize}
\item First order \begin{eqnarray}
\tilde{x}_{1}^{(1)}(\Omega ) & = & \frac{\sqrt{2}\xi _{x1}(\Omega )}{\left( i\Omega +1-\mu \right) }\, \, ,\nonumber \\
\tilde{y}_{1}^{(1)}(\Omega ) & = & \frac{\sqrt{2}\xi _{y1}(\Omega )}{\left( i\Omega +1+\mu \right) }\, \, ,\nonumber \\
\tilde{x}_{2}^{(1)}(\Omega ) & = & \frac{2\gamma _{r}\xi _{x2}(\Omega )}{\left( i\Omega +\gamma _{r}\right) }\, \, ,\nonumber \\
\tilde{y}_{2}^{(1)}(\Omega ) & = & \frac{2\gamma _{r}\xi _{y2}(\Omega )}{\left( i\Omega +\gamma _{r}\right) }\, \, .
\end{eqnarray}

\item Second order \begin{eqnarray}
\tilde{x}_{1}^{(2)}(\Omega ) & = & \frac{\left[ \tilde{x}_{1}^{(1)}\star \tilde{x}_{2}^{(1)}+\tilde{y}_{1}^{(1)}\star \tilde{y}_{2}^{(1)}\right] (\Omega )}{2\left( i\Omega +1-\mu \right) }\, \, ,\nonumber \\
\tilde{y}_{1}^{(2)}(\Omega ) & = & \frac{\left[ \tilde{x}_{1}^{(1)}\star \tilde{y}_{2}^{(1)}-\tilde{y}_{1}^{(1)}\star \tilde{x}_{2}^{(1)}\right] (\Omega )}{2\left( i\Omega +1+\mu \right) }\, \, ,\nonumber \\
\tilde{x}_{2}^{(2)}(\Omega ) & = & -\frac{\gamma _{r}\left[ \tilde{x}_{1}^{(1)}\star \tilde{x}_{1}^{(1)}-\tilde{y}_{1}^{(1)}\star \tilde{y}_{1}^{(1)}\right] (\Omega )}{2\left( i\Omega +\gamma _{r}\right) }\, \, ,\nonumber \\
\tilde{y}_{2}^{(2)}(\Omega ) & = & -\frac{\gamma _{r}\left[ \tilde{x}_{1}^{(1)}\star \tilde{y}_{1}^{(1)}\right] (\Omega )}{\left( i\Omega +\gamma _{r}\right) }\, \, .
\end{eqnarray}

\item Third order (sub-harmonic field) 
\end{itemize}
\begin{eqnarray}
\tilde{x}_{1}^{(3)}(\Omega ) & = & \frac{\left[ \tilde{x}_{1}^{(1)}\star \tilde{x}_{2}^{(2)}+\tilde{x}_{1}^{(2)}\star \tilde{x}_{2}^{(1)}+\tilde{y}_{1}^{(1)}\star \tilde{y}_{2}^{(2)}+\tilde{y}_{1}^{(2)}\star \tilde{y}_{2}^{(1)}\right] (\Omega )}{2\left( i\Omega +1-\mu \right) }\, \, ,\nonumber \\
\tilde{y}_{1}^{(3)}(\Omega ) & = & \frac{\left[ \tilde{x}_{1}^{(1)}\star \tilde{y}_{2}^{(2)}+\tilde{x}_{1}^{(2)}\star \tilde{y}_{2}^{(1)}-\tilde{y}_{1}^{(2)}\star \tilde{x}_{2}^{(1)}-\tilde{y}_{1}^{(1)}\star \tilde{x}_{2}^{(2)}\right] (\Omega )}{2\left( i\Omega +1+\mu \right) }\, \, .
\end{eqnarray}

\subsubsection{Squeezing Correlation spectrum}

The spectrum of the fields are given, for instance for the squeezed quadrature \( y_{1} \), by \begin{eqnarray}
\langle \tilde{y}_{1}(\Omega _{1})\tilde{y}_{1}(\Omega _{2})\rangle  & = & \langle \tilde{y}_{1}^{(1)}(\Omega _{1})\tilde{y}_{1}^{(1)}(\Omega _{2})\rangle +g^{2}\left\{ \langle \tilde{y}_{1}^{(2)}(\Omega _{1})\tilde{y}_{1}^{(2)}(\Omega _{2})\rangle \right. \nonumber \\
 & + & \left. \langle \tilde{y}_{1}^{(1)}(\Omega _{1})\tilde{y}_{1}^{(3)}(\Omega _{2})\rangle +\langle \tilde{y}_{1}^{(1)}(\Omega _{2})\tilde{y}_{1}^{(3)}(\Omega _{1})\rangle \right\} +\cdots 
\end{eqnarray}

The first order perturbation theory generates the usual linearized squeezed result as in quantum theory \begin{equation}
\langle \tilde{y}_{1}^{(1)}(\Omega _{1})\tilde{y}_{1}^{(1)}(\Omega _{2})\rangle =\frac{2\delta (\Omega _{1}+\Omega _{2})}{\Omega _{1}^{2}+(1+\mu )^{2}}\, \, ,
\end{equation}
 and, similarly, for the amplified fluctuation quadrature \begin{equation}
\langle \tilde{x}_{1}^{(1)}(\Omega _{1})\tilde{x}_{1}^{(1)}(\Omega _{2})\rangle =\frac{2\delta (\Omega _{1}+\Omega _{2})}{\Omega _{1}^{2}+(1-\mu )^{2}}\, \, ,
\end{equation}
 and for the pump quadratures, there is no first order squeezing \begin{equation}
\langle \tilde{x}_{2}^{(1)}(\Omega _{1})\tilde{x}_{2}^{(1)}(\Omega _{2})\rangle =\langle \tilde{y}_{2}^{(1)}(\Omega _{1})\tilde{y}_{2}^{(1)}(\Omega _{2})\rangle =\frac{4\gamma _{r}^{2}}{\Omega _{1}^{2}+\gamma _{r}^{2}}\delta (\Omega _{1}+\Omega _{2})\, \, .
\end{equation}

The next contribution to the squeezing field quadrature is: \begin{eqnarray}
\langle \tilde{y}_{1}^{(2)}(\Omega _{1})\tilde{y}_{1}^{(2)}(\Omega _{2})\rangle  & = & \frac{\gamma _{r}\delta (\Omega _{1}+\Omega _{2})}{\Omega _{1}^{2}+(1+\mu )^{2}}\left\{ \frac{1-\mu +\gamma _{r}}{(1-\mu )\left[ \Omega _{1}^{2}+(1-\mu +\gamma _{r})^{2}\right] }+\right. \nonumber \\
 & + & \left. \frac{1+\mu +\gamma _{r}}{(1+\mu )\left[ \Omega _{1}^{2}+(1+\mu +\gamma _{r})^{2}\right] }\right\} \, \, ,
\end{eqnarray}

and \begin{eqnarray}
\langle \tilde{y}_{1}^{(1)}(\Omega _{1})\tilde{y}_{1}^{(3)}(\Omega _{2})\rangle  & + & \langle \tilde{y}_{1}^{(1)}(\Omega _{2})\tilde{y}_{1}^{(3)}(\Omega _{1})\rangle =\frac{2\mu \gamma _{r}\delta (\Omega _{1}+\Omega _{2})}{\left[ \Omega _{1}^{2}+(1+\mu )^{2}\right] ^{2}}\nonumber \\
 & \times  & \left\{ -\frac{(1+\mu )(1-\mu +\gamma _{r})-\Omega _{1}^{2}}{(1-\mu )\left[ \Omega _{1}^{2}+(1-\mu +\gamma _{r})^{2}\right] }\right. \nonumber \\
 &  & \left. +\frac{(1+\mu )(1+\mu +\gamma _{r})-\Omega _{1}^{2}}{(1+\mu )\left[ \Omega _{1}^{2}+(1+\mu +\gamma _{r})^{2}\right] }+\frac{(1+\mu )}{\gamma _{r}(1-\mu ^{2})}\right\} \, \, .
\end{eqnarray}

\noindent and the internal (symmetrically ordered) squeezing spectrum is \begin{eqnarray}
S(\Omega ) & = & \frac{2}{\Omega ^{2}+(1+\mu )^{2}}+\frac{g^{2}\gamma _{r}}{\left[ \Omega ^{2}+(1+\mu )^{2}\right] ^{2}}\left\{ \frac{2\mu (1+\mu )}{\gamma _{r}(1-\mu ^{2})}\right. \nonumber \\
 & + & \left. \frac{(1-\mu +\gamma _{r})\Omega ^{2}+\left[ (1+\mu )^{2}+2\mu (1+\mu )\right] (1+\mu +\gamma _{r})}{(1+\mu )\left[ \Omega ^{2}+(1+\mu +\gamma _{r})^{2}\right] }\right. \nonumber \\
 & + & \left. \frac{(1+\mu +\gamma _{r})\Omega ^{2}+(1-\mu ^{2})(1-\mu +\gamma _{r})}{(1-\mu )\left[ \Omega ^{2}+(1-\mu +\gamma _{r})^{2}\right] }\right\} \, \, .
\end{eqnarray}

Of greater interest is the external squeezing spectrum, which is obtained by including both internal fields and the correlated reflected vacuum noise terms:

\begin{eqnarray}
V(\Omega ) & = & 1-\frac{4\mu }{\Omega ^{2}+(1+\mu )^{2}}+\frac{2g^{2}\gamma _{r}}{\left[ \Omega ^{2}+(1+\mu )^{2}\right] ^{2}}\left\{ \frac{\mu (1+\Omega ^{2}-\mu ^{2})}{\gamma _{r}(1-\mu ^{2})}\right. \nonumber \\
 & + & \left. \frac{\left[ (1-\mu )(1-\mu +\gamma _{r})-2\mu ^{2}\right] \Omega ^{2}+\left[1-\mu +\gamma _{r}\right]\left[ 1+\mu +\mu ^{2}+\mu ^{3}\right] }{(1-\mu )\left[ \Omega ^{2}+(1-\mu +\gamma _{r})^{2}\right] }\right. \nonumber \\
 & + & \left. \frac{\left[ (1+\mu )(1+\mu +\gamma _{r})+2\mu ^{2}\right] \Omega ^{2}+\left[1+\mu +\gamma _{r}\right]\left[ 1+3\mu +\mu ^{2}-\mu ^{3}\right] }{(1+\mu )\left[ \Omega ^{2}+(1+\mu +\gamma _{r})^{2}\right] }\right. \, \, .
\end{eqnarray}

\begin{figure}
{\centering \resizebox*{10cm}{10cm}{\includegraphics{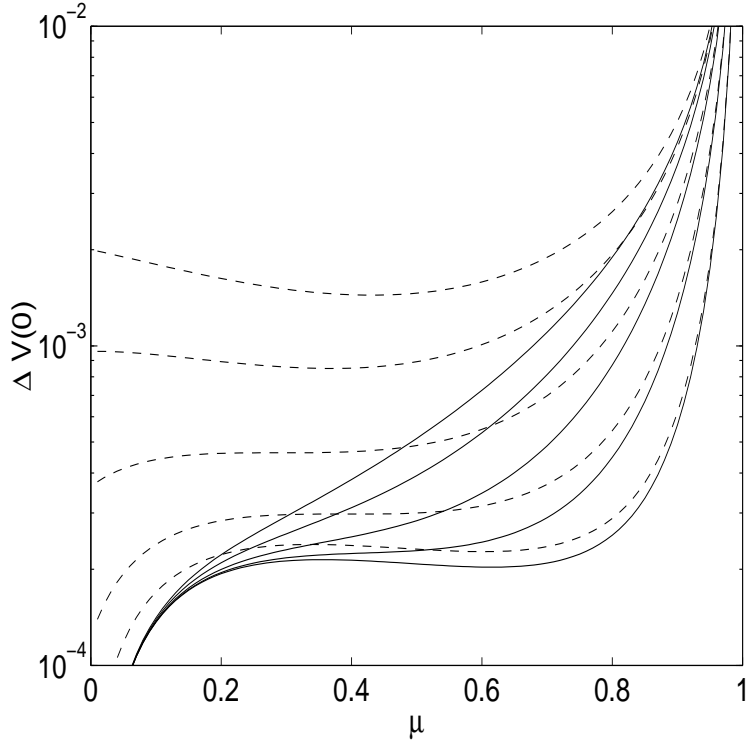}} \par}

\caption{Comparison of zero-frequency nonlinear squeezing spectrum between the positive- P (solid lines) and Wigner (dashed line) methods, with \protect\( g^{2}=.001\protect \). Values of \protect\( \gamma _{r}=.01,.1,1.,10.,100.\protect \) are used for the different lines plotted, with the lowest values of \protect\( \gamma _{r}\protect \) giving the smallest nonlinear correction. \label{NL2PWSPEC}}
\end{figure}

This semi-classical spectrum is quite different from positive-P calculation when \( \mu \rightarrow 0 \) but gives a compatible result near threshold, that is in the limit \( \mu \rightarrow 1 \). A detailed comparison of the zero-frequency behaviour is shown in Fig (\ref{NL2PWSPEC}).

This means that even when the pump is off, semiclassical theory gives a distorted vacuum spectrum due to the presence of the nonlinear crystal. This happens because in this theory the vacuum fluctuations are taken as real, and then two vacuum modes can interact inside the crystal as real fields. In the limit of \( \gamma _{r}\rightarrow 0 \) , the two spectra become compatible again, as the semiclassical theory decouples the second-harmonic mode from its vacuum input in this limit. In the case of threshold fluctuations, we can interpret the agreement as due to the large photon numbers involved - which means that the truncation approximation used for the semiclassical calculation is more reliable.

\subsection{Optimal Squeezing}

It is interesting to evaluate the squeezing or low-noise quantum correlations in the limit of zero frequency, that is in the resonance regime which is generally the frequency of maximum squeezing. We obtain from the positive-P result:

\begin{equation}
V(0)=1-\frac{4\mu }{(1+\mu )^{2}}+\frac{2\mu g^{2}}{(1+\mu )^{4}}\left[ 1+\frac{4\gamma _{r}\mu ^{2}(\gamma _{r}+2)}{(1-\mu )[(1+\gamma _{r})^{2}-\mu ^{2}]}\right] \, \, .
\end{equation}

Near threshold, where \( \mu \approx 1 \), we can set \( \mu =1+\delta  \), and expand in powers of \( \delta <0 \). Minimizing this result with respect to \( \delta  \), we find that, to leading order in \( g \), the optimal driving field is the solution to the following equation:\[
\delta ^{3}(2\delta +\gamma _{r}(\gamma _{r}+2))^{2}=g^{2}\gamma _{r}(\gamma _{r}+2)(4\delta -\gamma _{r}(\gamma _{r}+2))\]

This is a quintic equation, but it has simple closed form solutions in two limits, depending on whether \( \gamma _{r}\gg g^{2/3} \) or \( \gamma _{r}\ll g^{2/3} \). In the first case, the variance can be rewritten as:

\begin{equation}
V(0)=\frac{1}{4}\left[ \delta ^{2}+\frac{g^{2}}{2}-\frac{2g^{2}}{\delta }\right] \, \, .
\end{equation}
 Minimizing this result with respect to \( \delta  \) , we find that the minimum level of internal fluctuations occurs in a narrow frequency range near \( \Omega =0 \) , at a driving field just below threshold, with \( \delta =-g^{2/3} \) so that:

\begin{equation}
\mu _{opt}=1-g^{2/3}\, \, .
\end{equation}
 To leading order in \( g \) , the corresponding spectral variance is:

\begin{equation}
V_{opt}(0)=\frac{3}{4}g^{4/3}\, \, .
\end{equation}
 This result of an \( N^{-2/3} \) scaling confirms an approximate calculation of Plimak and Walls \cite{Plimak}, although the self-consistent method used by these authors makes it difficult to obtain the relevant driving field.

The physics of this is clearly that the onset of critical fluctuations starts to spoil the noise-reduction even before the critical point is reached at \( \mu =1 \). For example, with \( \gamma _{r}\approx 1 \) and \( \mu =0.9 \) , we find that \( V(0)\simeq 0.7\times 10^{-2} \) , or about \( 21dB \) below shot noise, as predicted from the analytic theory. This can also be seen from the way that the third order term includes contributions from the critical fluctuations in \( x_{1} \). A direct calculation from the full spectrum shows that this is a true minimum for all frequencies, even including \( \Omega >0 \) .

However, the situation clearly changes as \( \gamma _{r}\rightarrow 0 \) , in which case much greater levels of spectral noise-reduction are possible. This is plotted in Fig (\ref{OPT2DPSPEC}):
\begin{figure}
{\centering \resizebox*{10cm}{10cm}{\includegraphics{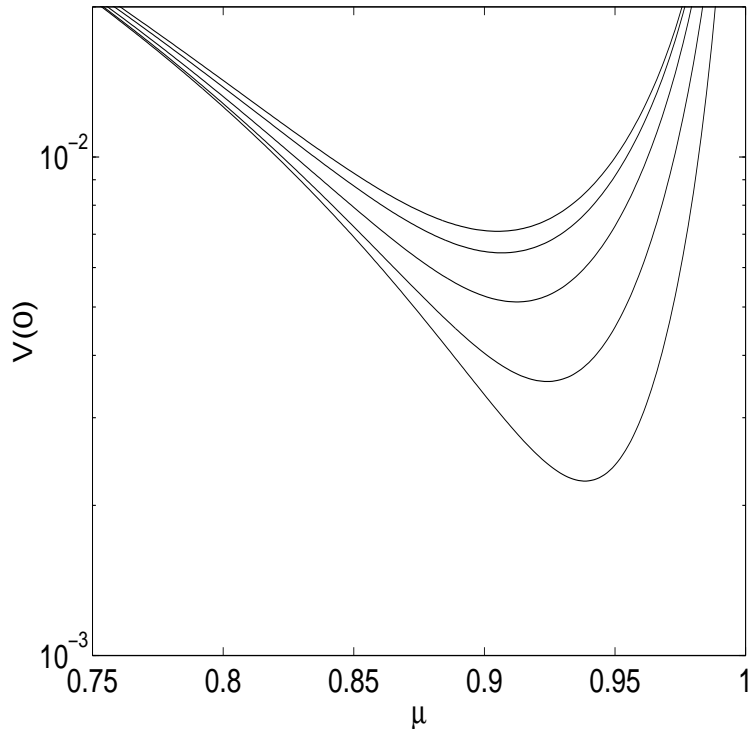}} \par}

\caption{Optimization of zero-frequency squeezing spectrum versus driving field using the positive-P method, with \protect\( g^{2}=.001\protect \). Values of \protect\( \gamma _{r}=0.001,.01,.1,1.,10.\protect \) are used for the different lines plotted, with the lowest values of \protect\( \gamma _{r}\protect \) giving the best results for squeezing. \label{OPT2DPSPEC}}
\end{figure}
Analytically, this limit gives the following result, provided that \( g^{2}\ll \gamma _{r}\ll g^{2/3} \) :

\begin{equation}
V(0)=\frac{1}{4}\left[ \delta ^{2}+\frac{g^{2}}{2}+\frac{2g^{2}\gamma _{r}}{\delta ^{2}}\right] \, \, .
\end{equation}
 Minimizing this result with respect to \( \delta  \) , we find that the minimum level of internal fluctuations occurs at a driving field very close to threshold, with:

\begin{equation}
\mu _{opt}=1-g^{1/2}(2\gamma _{r})^{1/4}\, \, .
\end{equation}
 The corresponding variance is therefore:

\begin{equation}
V_{opt}(0)=g\sqrt{\gamma _{r}/2}\ll g^{4/3}\, \, .
\end{equation}
 This result can be much smaller than predicted by the calculation of Plimak and Walls \cite{Plimak}, since the damping ratio can be reduced (at least in principle) to an arbitrarily low level - although still bounded below by \( g^{2} \), in order for perturbation theory to be applicable, so that we do not expect to obtain \( V_{opt}(0)<g^{2} \) . Of course, there are experimental limitations on this, due to absorption losses in the nonlinear medium at short wavelengths. Thus, for example, with the same value of \( g^{2}=0.001 \) as previously, but with \( \gamma _{r}=0.01 \) , we find that the minimum spectral noise is predicted to occur at a driving field of \( \mu =.93 \) , with a squeezing variance of \( 2.2\times 10^{-3} \) , or about \( 27dB \) below shot noise - about \( 6dB \) lower than before. 
\begin{figure}
{\centering \resizebox*{10cm}{10cm}{\includegraphics{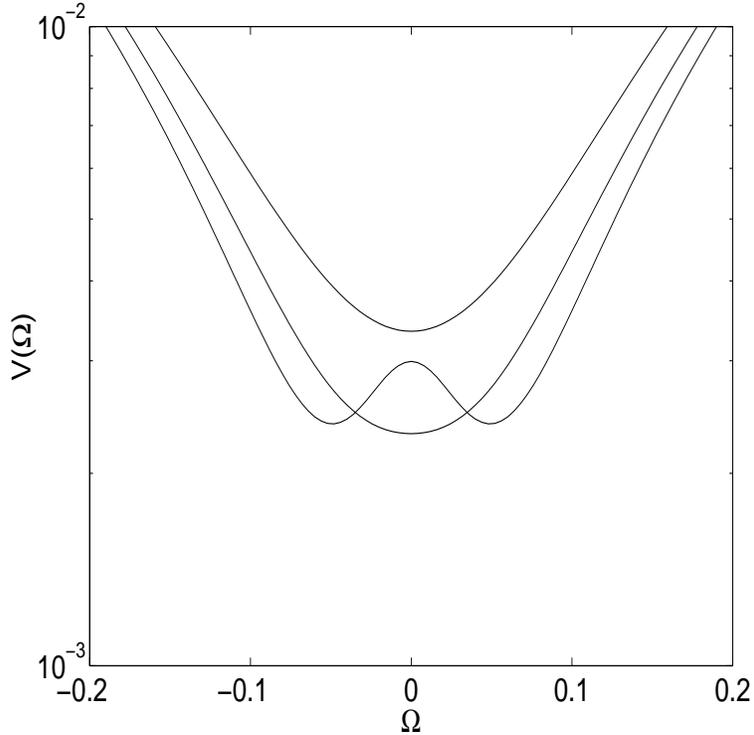}} \par}

\caption{Optimal zero-frequency squeezing spectrum versus frequency using the positive-P method, with \protect\( g^{2}=.001\protect \), \protect\( \gamma _{r}=0.01\protect \). Driving fields of \protect\( \mu =0.9,\, 0.93,\, 0.96\protect \) are used for the different lines plotted, with the higher driving fields giving the best results for squeezing, except at zero frequency. \label{TOT2dp01spec}}
\end{figure}
This operating regime also has the property that the optimum frequency of noise reduction moves away from zero frequency as the driving field is increased above the optimum value, towards threshold. At slightly higher driving fields than the optimum point, a bifurcation to a spectrum with two minima occurs, although with similar levels of noise reduction, as shown in Fig (\ref{TOT2dp01spec}). In this regime the results of the perturbation theory need to be checked by a full simulation of the stochastic equations. We have carried this out (see next section), and find that the full simulations do agree very well with the analytic predictions, even with this small damping ratio.

\subsection{Numerical Simulations}

The value of the nonlinear correction to the spectrum of the  squeezed quadrature \( V(\Omega ) \) can be worked out from a full numerical simulation\cite{Carmichael-SS-1986} of the relevant nonlinear stochastic equations. The optimal squeezing in the zero frequency part of the squeezing spectrum is predicted to scale as \( N^{-2/3} \) with roughly equal values of decay rates. For the simulations, we chose values of \( N=g^{-2}=10^{3} \), \( \gamma _{r}=0.5 \). The simulations used a total dimensionless time-interval of \( \tau _{max}=1000 \). To ensure equilibrium, only the last \( 500 \) time units were utilized in the Fourier transforms. Time steps of \( \Delta \tau =0.1 \) and \( \Delta \tau =0.2 \) were compared to ensure convergence. The algorithmic technique is described elsewhere \cite{Drummond-1991}, and uses a semi-implicit central partial difference technique. To obtain the small nonlinear corrections near the optimum squeezing, we simulated the difference between the linear and nonlinear forms of the stochastic equation, in order to minimize sampling errors. It was also useful to initialize the \( x \) quadratures with a Gaussian ensemble close to the known steady-state variance, in order to reduce the time taken to achieve equilibrium. Typically, the relative error in the correlations due to finite step-size was around \( 10^{-4} \) with these step-sizes.

For these parameters the optimal driving field is predicted to occur at \( \mu =0.9 \), or approximately \( 80\% \) of the critical intensity. We used \( 10^{5} \) trajectories to improve the relative error due to sampling with a finite trajectory population, giving relative sampling errors of less than \( 10^{-2}. \)
\begin{figure}
{\centering \resizebox*{10cm}{10cm}{\includegraphics{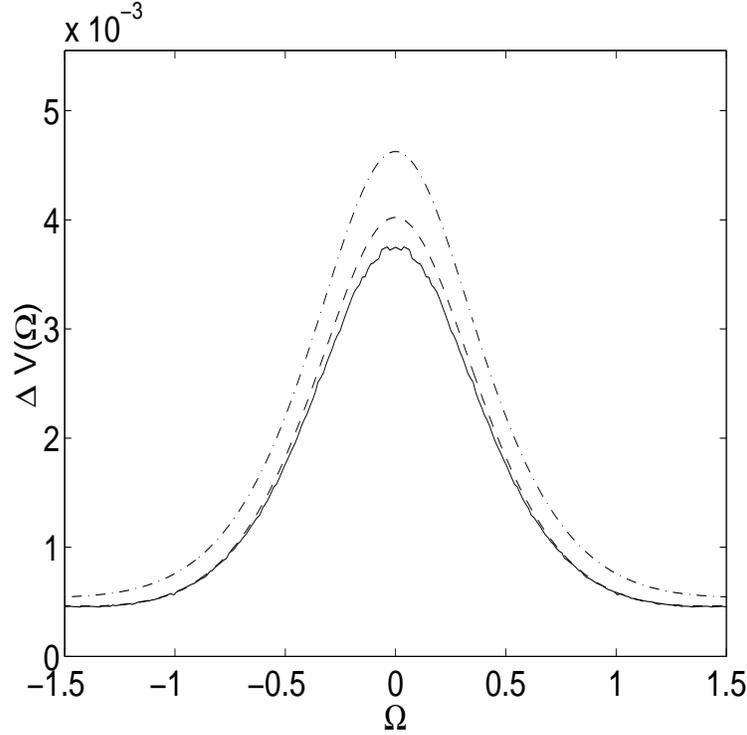} }\par}

\caption{Numerically simulated optimum nonlinear squeezing with \protect\( g^{2}=.001,\protect \) \protect\( \gamma _{r}=0.5\protect \), \protect\( \mu =0.9\protect \). Solid line is the +P simulation result, dashed line the analytic prediction from perturbation theory, dashed-dotted line the Wigner prediction.\label{OTSIM9}}
\end{figure}

The calculated squeezing moment from the SDE simulations was: \( \langle \widehat{y}^{2}_{1}\rangle +0.5=.0271\pm 10^{-4} \). This is in excellent agreement with the below-threshold expansion, which gives \( \langle Y^{2}_{1}\rangle +0.5=.0272 \), as this is just outside the critical region.

We find that the spectral predictions are also well verified by the simulations. These resulted in a value for the nonlinear correction to the zero-frequency spectrum, of \( \Delta V(0)=V(0)-V^{(1)}(0)=3.75\times 10^{-3}\pm .02\times 10^{-3} \). By comparison, the analytic theory, worked to fourth order in \( g \), gives the prediction that \( \Delta V(0)=4.02\times 10^{-3} \) . The residual difference of about \( 5\% \) - which is significant compared to sampling error - can be attributed to the fact that there are higher order corrections that are not included in the analytic theory, and these are more significant in the zero-frequency spectrum than they are in the moment calculation. Fig.~(\ref{OTSIM9}) shows the detailed results of the simulation. 
\begin{figure}
{\centering \resizebox*{10cm}{10cm}{\includegraphics{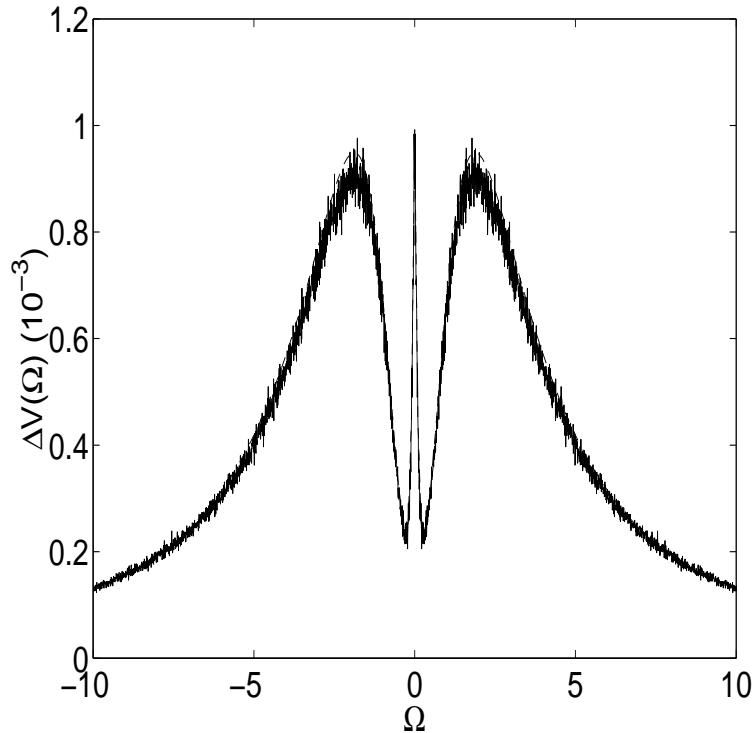}} \par}

\caption{Numerically simulated optimum nonlinear squeezing with \protect\( g^{2}=.001\protect \), \protect\( \gamma _{r}=0.01\protect \), \protect\( \mu =0.93\protect \). Solid line is the +P simulation result, dashed line the analytic prediction from +P perturbation theory, ( results are identical with the Wigner prediction). Analytic predictions in this case are very close to numerical simulation results. }

\label{OPTSIM01}
\end{figure}

In the analytic theory, we found that a smaller decay rate for the second-harmonic is predicted to yield a better squeezing optimum, as a function of driving field. In Fig. (\ref{OPTSIM01}) we verify this to be the case, by carrying out a full numerical simulation for \protect\( \gamma _{r}=0.01\protect \), \protect\( \mu =0.93\protect \), with \( \tau _{max}=2000 \), time steps of \( \Delta \tau =0.05 \) and \( \Delta \tau =0.1 \) for error-checking,  and \( 10^4 \) trajectories. The results show that the analytic predictions and numerical simulations are almost indistinguishable in this regime. The sampling error was relatively larger, possibly due to the fact that the absolute noise levels are lower here. The agreement indicates that the perturbation theory is an excellent approximation to the full nonlinear equations with these parameters.

\section{Conclusion}

We have calculated the nonlinear quantum fluctuations in a parametric oscillator below the classical threshold, using a nonlinear stochastic positive-P theory, with both asymptotic approximations and a numerical technique. There is excellent agreement between numerical and analytic calculations. Corresponding results for the Keldysh diagram method require a summation over infinite sets of diagrams, in order to fully include the reservoirs. The advantage of the present method is due to the fact that the coherent state basis is a more natural basis set for an open system, since it allows the damping reservoirs to be treated non-perturbatively.

Optimal squeezing in the output spectra corresponding to these moments were estimated. We found that the best squeezing in the zero frequency part of the squeezing spectrum scales like \( N^{-2/3} \) just below threshold, \emph{provided the two fields have similar damping rates.} In other words, at the true critical threshold - where the linear squeezing is optimized - the nonlinear corrections are too large to give the lowest overall zero-frequency squeezing. Instead, one should operate below the critical point to optimize the spectral squeezing. Using an entirely different method, a calculation by Plimak and Walls \cite{Plimak} also predicted that the optimum zero frequency squeezing spectrum scales like \( N^{-2/3} \) , or equivalently, as \( I^{-2/3} \) for a given input flux \( I \) . Our general scaling results agree with theirs, except with a different spectrum. We attribute the difference to the systematic +P stochastic diagram procedure used here to calculate the spectrum, rather than the Feynman diagram method - which involves additional approximations. 

We also found a new regime in which the lower limit to the spectral noise-reduction depends on the decay rate of the second-harmonic field, which can be reduced to an arbitrarily low level. This has a reasonable physical interpretation, since the second-harmonic losses are essentially parasitic losses, which do not contribute to the desired squeezing output. The ultimate limit to squeezing in this regime is set by even higher order terms in perturbation theory. We conjecture that optimization of both the driving field and the relative decay rate may result in a final squeezing variance scaling as \( \gamma_{1}/I_{c} \) .

A calculation with the truncated Wigner method, or semiclassical technique, was also carried out. Well below threshold, we found that while the linear terms agreed with full quantum calculation, nonlinear corrections and higher order correlations tended to disagree, especially for high second-harmonic losses. However, near the critical point, the situation changed. Here, even though the dominant terms are nonlinear, we found excellent agreement between the two methods.

\begin{acknowledgements}

We acknowledge the financial support of FAPESP (Brazil) and the Australian Research
Council. One of the authors (K.D.) would like to acknowledge the hospitality
of the University of Queensland.
\end{acknowledgements}

\end{document}